\documentclass[aps,prl,reprint,superscriptaddress,footinbib,longbibliography,floatfix]{revtex4-2}
\usepackage{graphicx}
\usepackage{dcolumn}
\usepackage{bm}
\usepackage{booktabs}
\usepackage{amsmath}
\usepackage{amssymb}
\usepackage{mathtools}
\usepackage{dsfont}
\usepackage{float}
\usepackage{pgfplots}
\usepgfplotslibrary{groupplots}
\pgfplotsset{compat=1.18}
\usepackage[colorlinks=true,breaklinks=true,allcolors=blue]{hyperref}
\def\Ll{\ensuremath{{\cal L}}}
\def\Zz{\ensuremath{{\cal Z}}}
\def\Tr{\ensuremath{{\operatorname{Tr}}}}
\newcommand{\ket}[1]{\lvert #1\rangle}
\newtheorem{theorem}{Theorem}
\newtheorem{proposition}{Proposition}

\begin{document}
	
	\title{Krylov Edge Spectroscopy of Symmetry-Protected Topological Phases}
	
	\author{Heiko Georg Menzler}
	\email{heiko.menzler@uni-goettingen.de}
	\affiliation{Institute for Theoretical Physics, Georg-August-Universit\"at G\"ottingen, Friedrich-Hund-Platz 1, 37077 G\"ottingen, Germany}
	
	\author{Rishabh Jha}
	\email{rishabh.jha@usc.edu}
	\affiliation{Department of Physics and Astronomy, University of Southern California, Los Angeles, California 90089-0484, USA}
	
	\begin{abstract}
		We introduce \textit{Krylov edge spectroscopy}, a many-body operator-space protocol for detecting and classifying one-dimensional symmetry-protected topological phases from local boundary dynamics. A Hermitian boundary operator generates a semi-infinite Krylov hopping chain whose boundary weight obeys an exact zero-frequency normalizability criterion. Open-periodic, boundary-bulk, and symmetry-preserving boundary-perturbation tests identify protected boundary memory, while a finite-depth leakage residual certifies when an explicitly reconstructed operator is already near zero frequency. Classification minimizes the normalized commutator over symmetry-resolved boundary operators in fixed charge sectors. For bosonic $\mathbb{Z}_N \times \mathbb{Z}_N$ phases, the recovered endpoint charge gives the cohomology label. The method requires no explicit ground-state wavefunction, entanglement spectrum, or guessed dressed edge or string operator. For finite-range Hamiltonians, locality eliminates full-system exact diagonalization and guarantees thermodynamic-limit convergence at fixed Krylov depth. We demonstrate the protocol in cluster, clock, Haldane, and trivial spin-1 chains. In the exactly solvable cluster chain, the transition between the gapped topological and trivial phases manifests as a localization-delocalization transition of the Krylov edge mode on the Krylov chain. This transition occurs precisely at the bulk gap closing, and its localization-length exponent coincides with the Ising correlation-length exponent. Through operator Krylov dynamics, our work turns local boundary evolution into a direct spectroscopy of many-body topology.
	\end{abstract}
	
	\maketitle
	
	\textit{Introduction.---}%
	Symmetry-protected topological (SPT) phases preserve their protecting symmetry and have short-ranged bulk correlations, yet they cannot be connected to a trivial product state without closing the bulk gap or breaking that symmetry~\cite{Hasan2010,Qi2011,Senthil2015,Gu2009,Chen2011prb,Chen2013prb,Chen2013science,Schuch2011,Levin2012}. In one dimension an open boundary exposes this obstruction: the effective endpoint actions of symmetries that commute in the bulk can acquire a phase when exchanged~\cite{Chen2011prb,Pollmann2010,Pollmann2012,Else2014}. In the spin-1 Haldane chain, two half-turn spin rotations commute globally but anticommute on the effective spin one-half at either edge. For two cyclic symmetries, this exchange phase labels the bosonic SPT class~\cite{Chen2011prb,Chen2013prb,Chen2013science,Schuch2011}.
	
	This edge relation is physical but often hidden from local static probes. Representative systems include Su--Schrieffer--Heeger~\cite{SSH1979,HangHaasJha2026EdgeMemory}, Haldane and AKLT~\cite{Haldane1983,AKLT1987,Kennedy1992Jan,Kennedy1992Jul,Pollmann2010,Pollmann2012}, cluster~\cite{Raussendorf2001,Son2011}, and clock chains~\cite{Fendley2012}. Entanglement spectra examine a reduced state~\cite{Li2008,Pollmann2010,GadgePremJha2026TET}, while string and symmetry-based diagnostics require a chosen nonlocal operator or a prior description of how symmetry divides between the edges~\cite{denNijs1989,Kennedy1992Jan,Kennedy1992Jul,Pollmann2012}. We ask whether local boundary dynamics can reveal protected memory and reconstruct the edge symmetry action without either input.
	
	Our organizing idea turns repeated commutators of a local boundary observable with the Hamiltonian into a one-dimensional chain of orthogonal observables; the Lanczos recursion performs the orthogonalization. A normalizable end state gives persistent boundary memory. Open-versus-periodic, boundary-versus-bulk, and symmetry-preserving boundary-perturbation comparisons establish whether that memory is boundary selective and protected. Classification then searches a growing boundary region for the slowest operator in each symmetry sector and reads the phase from the recovered endpoint charge and algebra.
	
	The practical inputs are the Hamiltonian, the onsite symmetry matrices, a chosen inverse temperature $\beta$, and a simple boundary observable for detection; the classification search space is generated from the onsite symmetry action. Krylov methods already describe observable spreading, scrambling, and localization~\cite{Parker2019,BAIGUERA20261,NANDY20251,rabinovici2025krylovcomplexity,YatesAbanovMitra2020PRL,YatesAbanovMitra2020PRB,YatesMitra2021,vonKeyserlingk2018,Balasubramanian2022,Bhattacharyya2023Oct,MenzlerJha2024,AlishahihaVasli2026MemoryCores}. Quadratic Kitaev chains provide an exactly solvable single-particle limit~\cite{Kitaev2001,JhaMenzler2026lr}, while interactions can change the one-dimensional fermionic classification~\cite{FidkowskiKitaev2011}. Earlier work mapped strong and almost-strong edge modes to SSH-like Krylov chains~\cite{YatesAbanovMitra2020PRL}. We establish an exact zero-frequency normalizability criterion with ordered thermodynamic and Krylov-depth limits at fixed finite $\beta$, add geometry and symmetry-preserving boundary controls, and recover the projective/cohomology label from symmetry-resolved endpoint operators.

	\textit{Krylov chain and edge memory.---}
	Throughout, $H$ is a finite-range one-dimensional Hamiltonian and we use the Gibbs state $\rho_\beta=e^{-\beta H}/\Tr e^{-\beta H}$. Because each protecting symmetry commutes with $H$, $\rho_\beta$ automatically has the same symmetry. The solvable cluster benchmark and the fixed-point clock benchmark~\footnote{Here ``fixed point'' means an exactly solvable representative of the SPT phase built from commuting local terms, with strictly local edge operators before perturbations; it is not a critical fixed point. For the clock model these edge operators commute exactly with $H$, so $\beta=0$ already resolves their zero-frequency sector.} are already resolved at $\beta=0$; finite $\beta$ remains allowed but is not needed for these benchmarks. Away from such solvable limits, larger $\beta$ emphasizes low-energy matrix elements and can sharpen edge resolution. Temperature enters only through the operator-space metric below; here $\beta$ sets spectral resolution rather than a thermodynamic finite-temperature SPT order parameter.
	
	For any operator $A$, define $\delta A=A-\Tr(\rho_\beta A)\mathds{1}$ and
	\begin{equation}
		(A|B)_S=\tfrac{1}{2}\Tr[\rho_\beta\{\delta A^\dagger,\delta B\}].
		\label{eq:inner}
	\end{equation}
	Let the onsite symmetry generators be $U_i=\bigotimes_j u_{i,j}$. To generate unbiased one-site initial operators, decompose a complete traceless Hermitian basis on the boundary site under $O\mapsto u_{i,0}Ou_{i,0}^\dagger$, discard zero and linearly dependent outputs, and retain every independent direction. If a protected edge operator has a one-site component, completeness guarantees that at least one resulting initial operator overlaps it and can satisfy the detection criterion. If all one-site probes fail, test complete bases on larger boundary regions and assign triviality only if every probe continues to fail once the support is large enough to capture the edge dressing while remaining localized near one boundary. Our minimal benchmark choices are $Z_0$ for the cluster chain, the two Hermitian quadratures $(Z_0+Z_0^\dagger)/\sqrt2$ for an $N>2$ clock, and $\{S_0^x,S_0^y,S_0^z\}$ for spin-1. Automatic generation and symmetry-sector decomposition are given in the SM~\cite{SM}.
	
	Run Lanczos for each resulting Hermitian initial operator $O_0$, normalized by $(O_0|O_0)_S=1$. With $\Ll=[H,\cdot]$, Lanczos produces orthonormal Krylov operators $O_n$ and hoppings $b_n\geq0$ until the recursion terminates. If $H\ket{m}=E_m\ket{m}$, $\rho_\beta\ket{m}=r_m\ket{m}$, and $O_{mn}=\langle m|\delta O_0|n\rangle$, define the frequency weight of $O_0$ by
	\begin{equation}
		d\mu_{O_0}(\omega)=\tfrac12\sum_{m,n}(r_m+r_n)|O_{mn}|^2\delta(\omega-E_m+E_n)d\omega.
		\label{eq:frequency_measure}
	\end{equation}
	Its Fourier transform is the connected autocorrelation $C_{O_0}(t)=(O_0|e^{it\Ll}O_0)_S=\int e^{it\omega}d\mu_{O_0}(\omega)$. Hermiticity makes the weight even in $\omega$, so the diagonal Lanczos coefficients vanish. The Krylov problem is therefore a nearest-neighbor tight-binding chain in operator space, with site $n$ representing $O_n$ and hopping $b_n$ between sites $n-1$ and $n$. On sites $n=0,\ldots,2K$ its matrix is
	\begin{equation}
		T_K=\begin{pmatrix}
			0&b_1&&\\
			b_1&0&b_2&\\
			&b_2&0&\ddots\\
			&&\ddots&\ddots
		\end{pmatrix}.
		\label{eq:krylov_matrix}
	\end{equation}
	A finite many-body system terminates when a hopping vanishes. Before that termination, the zero-energy wavefunction of this Krylov chain obeys
	\begin{equation}
		\psi_{2m+1}=0,\qquad \psi_{2m}=(-1)^m\prod_{j=1}^{m}\frac{b_{2j-1}}{b_{2j}}\psi_0.
		\label{eq:zero_recur}
	\end{equation}
	The weight at the boundary site $n=0$ is
	\begin{equation}
		\Zz_K^{-1}=1+\sum_{m=1}^{K}\prod_{j=1}^{m}\left(\frac{b_{2j-1}}{b_{2j}}\right)^2.
		\label{eq:intro_ZK}
	\end{equation}
	Since $\mu_{O_0}(\{0\})$ is exactly the nondecaying zero-frequency weight of the initial operator, the infinite-depth statement is the following.

	\begin{theorem}[Normalizability criterion]
		For the semi-infinite Krylov chain defined by the limiting hoppings $b_n$, assume that the infinite tridiagonal operator has a unique self-adjoint realization. Then $\Zz_\infty\equiv\mu_{O_0}(\{0\})>0$ precisely when the Krylov zero mode is normalizable:
		\begin{equation}
			\Zz_K\to\Zz_\infty>0 \quad				\Longleftrightarrow \quad
			\sum_{m\geq1}\prod_{j=1}^{m}\left(\frac{b_{2j-1}}{b_{2j}}\right)^2<\infty.
			\label{eq:normalizable}
		\end{equation}
	\end{theorem}
	The limiting weight $\Zz_\infty$ is the stationary correlation value of Ref.~\cite{Gamayun2025Jul}. The balanced Kitaev chain is the single-particle subset of this construction: its operator recursion closes on Majoranas and becomes an alternating hopping chain (see \cite{JhaMenzler2026lr}). The explicit reduction and reconstructed Majorana edge mode are given in the SM~\cite{SM}; interactions restore the full many-body operator problem. 
	
	Locality fixes the system-size limit before the depth limit. Write ${\rm ad}_H^j(O_0)=\Ll^jO_0$ for the $j$-fold nested commutator and let $\Lambda_n$ contain every physical site reached for $0\leq j\leq n$. The Krylov Gram matrix through depth $2K$ is $G_{mn}=(\Ll^mO_0|\Ll^nO_0)_S$ for $0\leq m,n\leq2K$.
	
	\begin{proposition}[Fixed-depth size limit]
		\label{prop:fixed_depth}
		Fix $K$ while increasing $L$ using the same finite-range one-dimensional local interactions, with either open or periodic boundaries. The couplings near any fixed region are independent of $L$ and repeat in the bulk with a fixed finite spatial period, such as period two for the clock chain. Write $\Delta_Lx=x(L)-x(\infty)$ and assume that $G$ has a nonsingular thermodynamic limit. Once the support generated through depth $2K$ remains local, meaning that $\Lambda_{2K}$ has neither reached the opposite end of an open chain nor extended around a periodic ring, then for $1\leq n\leq2K$,
		\begin{subequations}\label{eq:fixed_depth_limit}
			\begin{align}
				\Delta_L b_n&=\Delta_L\Zz_K=0,\qquad \beta=0,
				\label{eq:fixed_depth_beta0}\\
				\Delta_L b_n&=O(e^{-\gamma_\beta L}),\qquad 0<\beta<\infty,
				\label{eq:fixed_depth_bn}\\
				\Delta_L\Zz_K&=O(e^{-\gamma_\beta L}),\qquad 0<\beta<\infty,
				\label{eq:fixed_depth_Z}
			\end{align}
		\end{subequations}
		where $\gamma_\beta>0$ is the finite-temperature, system-size independent, boundary-decay rate at fixed $K$.
	\end{proposition}
	Equation~\eqref{eq:fixed_depth_beta0} follows from size consistency, finite commutator support, and exact trace factorization. At $0<\beta<\infty$, exponential locality of one-dimensional Gibbs states makes local observables exponentially insensitive to a remote boundary and to the distant bond closing a periodic chain~\cite{Araki1969Gibbs,Capel2025Gibbs}; nonsingularity of the finite Krylov problem then gives Eqs.~\eqref{eq:fixed_depth_bn} and \eqref{eq:fixed_depth_Z}~\footnote{Here $f(L)=O(e^{-\gamma L})$ means $|f(L)|\leq Ce^{-\gamma L}$ for all sufficiently large $L$, with $C$ and $\gamma$ independent of $L$ at fixed $K$.}. An exactly terminated recursion is already size independent and stops at its last nonzero Krylov vector. All models considered in this work, namely the cluster, clock, AKLT, and large-$D$ chains at fixed local dimension, satisfy the finite-range and finite-dimensional hypotheses. Proofs are given in the SM~\cite{SM}.
	
	The controlled order is therefore $L\to\infty$ at fixed $K$, followed by the $K$-flow. If the sum in Eq.~\eqref{eq:normalizable} diverges, odd truncations still possess a finite-chain zero mode but its boundary weight vanishes. For example, if $b_n=1$, every product in Eq.~\eqref{eq:intro_ZK} equals unity, so $\Zz_K^{-1}=1+K$ and therefore $\Zz_K=1/(K+1)$. Thus a finite plateau is evidence, while a positive infinite-depth limit is the necessary-and-sufficient zero-frequency statement. The condition $b_1=0$ alone is inconclusive because an ordinary conserved or symmetry-forbidden local operator can also have zero first hopping.
	
	The same Krylov data reconstruct the corresponding slow operator. Define $\alpha_0=1$ and $\alpha_m=\prod_{j=1}^{m}b_{2j-1}/b_{2j}$.
	
	\begin{proposition}[Finite-depth leakage]
		For any $K$ for which $O_0,\ldots,O_{2K}$ exist, define
		\begin{align}
			A_K&=\sqrt{\Zz_K}\sum_{m=0}^{K}(-1)^m\alpha_mO_{2m},\nonumber\\
			\varepsilon_K&\equiv\|[H,A_K]\|_S=b_{2K+1}|\alpha_K|\sqrt{\Zz_K}.
			\label{eq:leakage}
		\end{align}
		If $b_{2K+1}>0$, then $\Ll A_K=(-1)^Kb_{2K+1}\alpha_K\sqrt{\Zz_K}\,O_{2K+1}$. If $b_{2K+1}=0$, the recursion terminates there and $A_K$ is an exact zero-frequency operator with $\varepsilon_K=0$.
	\end{proposition}
	The cancellations are proved in the SM~\cite{SM}. For every $\Omega>0$, Eq.~\eqref{eq:leakage} implies $\mu_{A_K}(|\omega|\geq\Omega)\leq\varepsilon_K^2/\Omega^2$ and, for Hermitian $A_K$, $C_{A_K}(t)\geq1-t^2\varepsilon_K^2/2$. Thus $\varepsilon_K$ has a direct physical meaning: it bounds how much of this explicitly reconstructed operator lies away from zero frequency.
	
	Using the same $\beta$, initial boundary operator $O_0$, and Krylov depth $K$ in each geometry, define
	\begin{equation}
		R_K=\frac{\Zz_K^{\rm OBC}}{\Zz_K^{\rm PBC}},\qquad B_K=\frac{\Zz_K^{\rm boundary}}{\Zz_K^{\rm bulk}}.
		\label{eq:ratios}
	\end{equation}
	A protected signal has sizable, depth-stable $\Zz_K^{\rm OBC}$, $R_K,B_K\gg1$, and stable size flow. The two ratios answer different questions: $R_K$ removes the physical boundary, while $B_K$ moves the initial operator into the bulk. A global conserved quantity typically gives $R_K=O(1)$.
	
	An accidentally free endpoint is tested by a symmetry-preserving boundary perturbation
	\begin{align}
		H(h_\partial)&=H+h_\partial V_\partial,\nonumber\\
		[V_\partial,U_i]&=0\qquad\text{for every generator }U_i.
		\label{eq:pinning_general}
	\end{align}
	Here $h_\partial$ is real and $V_\partial$ is supported near the ends. Because both $H$ and $V_\partial$ are local, symmetry preservation can be checked using only the onsite matrices $u_{i,j}$ on the support of each term. A projective SPT edge can dress or move under such a perturbation while its protected boundary sector remains; an accidental free endpoint can be removed. The same boundary perturbations are used as robustness checks after classification below.
	
	The four benchmark Hamiltonians are introduced here once and used throughout. We label every chain by $j=0,\ldots,L-1$. For the clock model, let $\mathcal I_{\rm O}=\{1,\ldots,L-2\}$ under OBC and $\mathcal I_{\rm P}=\{0,\ldots,L-1\}$ under PBC, with periodic indices understood modulo even $L$. Then
	\begin{align}
		H_{\rm cl}(\lambda)&=-\sum_{j=1}^{L-2}Z_{j-1}X_jZ_{j+1}-\lambda\sum_{j=0}^{L-1}X_j,
		\label{eq:Hcluster}\\
		K_j^{(p)}&=Z_{j-1}^{\epsilon_jp}X_jZ_{j+1}^{-\epsilon_jp},\qquad \epsilon_j=(-1)^{j+1},\nonumber\\
		H_p^\nu&=-\tfrac12\sum_{j\in\mathcal I_\nu}(K_j^{(p)}+K_j^{(p)\dagger}),\qquad p\in\mathbb Z_N,
		\label{eq:clockH}\\
		H_{\rm AKLT}&=\sum_{j=0}^{L-2}\left[\mathbf S_j\cdot\mathbf S_{j+1}+\tfrac13(\mathbf S_j\cdot\mathbf S_{j+1})^2\right],
		\label{eq:HAKLT}\\
		H_{\rm large-D}&=H_{\rm AKLT}+D\sum_{j=0}^{L-1}(S_j^z)^2,\qquad D=3.
		\label{eq:HlargeD}
	\end{align}
	The cluster generators are $U_{\rm e}=\prod_{j\ {\rm even}}X_j$ and $U_{\rm o}=\prod_{j\ {\rm odd}}X_j$; the clock generators are $U_1=\prod_{j\ {\rm even}}X_j$ and $U_2=\prod_{j\ {\rm odd}}X_j$. The spin-1 pair is $U_\alpha=\prod_j u_\alpha$ with $u_\alpha=e^{i\pi S^\alpha}$ for $\alpha=x,z$. In the basis $\{\ket{1},\ket{0},\ket{-1}\}$, $u_z={\rm diag}(-1,1,-1)$ and $u_x=\left(\begin{smallmatrix}0&0&-1\\0&-1&0\\-1&0&0\end{smallmatrix}\right)$. The corresponding boundary perturbations are
	\begin{align}
		V_\partial^{\rm cl}&=X_0+X_{L-1},\nonumber\\
		V_\partial^{\rm clock}&=\tfrac12(X_0+X_0^\dagger+X_{L-1}+X_{L-1}^\dagger),\nonumber\\
		V_\partial^{\rm spin}&=\sum_{e\in\{0,L-1\}}[(S_e^z)^2-2\mathds1/3].
		\label{eq:boundary_perturbations}
	\end{align}
	Each expression is Hermitian, boundary supported, and invariant under the stated onsite symmetries. Systematic generation of such perturbations from the local symmetry matrices is given in the SM~\cite{SM}.

	The solvable cluster and fixed-point clock benchmarks use $\beta=0$, while finite $\beta$ resolves the low-energy edge sector in the spin-1 comparison. A nonzero first hopping $b_1$, as in AKLT, shows that the microscopic initial operator is not itself conserved. Under the $D_2$-preserving perturbations in Eq.~\eqref{eq:boundary_perturbations}, $b_1$ can increase while $\Zz_K^{\rm OBC}$ remains sizable, showing that the protected boundary memory survives through dressing of the edge operator rather than conservation of the microscopic probe. The symmetry checks and boundary-perturbation analysis are given in the SM~\cite{SM}.
	
	\textit{Krylov recovery of the edge operator.---}%
	Detection asks whether a protected boundary sector is present. Classification reconstructs the symmetry action carried by that sector. For global onsite generators $U_i=\bigotimes_j u_{i,j}$, the low-energy action of an open SPT chain factorizes as
	\begin{equation}
		U_i\simeq V_i^{\rm L}V_i^{\rm R},
		\label{eq:fractionalization}
	\end{equation}
	up to local dressing and exponentially small finite-size corrections. The SPT label is the algebra of the left endpoints $V_i^{\rm L}$~\cite{Chen2011prb,Pollmann2012,Else2014}.
	
	Choose a boundary window of $\ell$ sites and a local operator basis $\{O_a\}$ supported there. Define
	\begin{equation}
		C_{ab}=(O_a|O_b)_S,\qquad D_{ab}=([H,O_a]|[H,O_b])_S.
		\label{eq:CD}
	\end{equation}
	For $O(v)=\sum_av_aO_a$, define the normalized commutator stiffness
	\begin{equation}
		\kappa(v)\equiv
		\frac{\|[H,O(v)]\|_S^2}{\|O(v)\|_S^2}
		=\frac{v^\dagger Dv}{v^\dagger Cv}.
		\label{eq:rayleigh}
	\end{equation}
	After removing null directions of $C$, minimizing $\kappa(v)$ over the boundary window gives
	\begin{equation}
		Dv=\kappa Cv.
		\label{eq:generalized}
	\end{equation}
	For normalized $A$, $\kappa(A)=\|[H,A]\|_S^2$. When $A$ is Hermitian, $\kappa(A)=b_1^2$; for a general charged operator, the classification uses $\kappa(A)$ directly.
	The SM~\cite{SM} proves the direct bounds $\mu_A(|\omega|\geq\Omega)\leq\kappa(A)/\Omega^2$ and $C_A(t)\geq1-t^2\kappa(A)/2$ for Hermitian $A$. Minimizing a local commutator norm has been used to identify slow local operators~\cite{KimBanulsCiracHastingsHuse2015}; here the same minimization is performed in symmetry-resolved boundary spaces and becomes an SPT classifier.
	For finite-range $H$, the matrices $C$ and $D$ need only Gibbs marginals on the search window and its interaction collar. No full-chain diagonalization is required: at $\beta=0$ they are exact local normalized traces, while at finite $\beta$ a guarded local Gibbs calculation converges as its guard grows; see the SM~\cite{SM}.
	
	For a symmetry sector $\bm q$, let $\kappa_{1,\bm q}^{\rm OBC/PBC}(\ell)$ be the lowest stiffness at boundary size $\ell$ and define
	\begin{equation}
		\eta_{\bm q}(\ell)=\frac{\kappa_{1,\bm q}^{\rm PBC}(\ell)}{\kappa_{1,\bm q}^{\rm OBC}(\ell)}.
		\label{eq:eta}
	\end{equation}
	A clean edge branch has small OBC stiffness, a large OBC-PBC contrast, and a stable charge and recovered operator as $\ell$ grows. In practice, increase $\ell$ while tracking the same branch. If $\eta_{\bm q}$ drops sharply from $\ell$ to $\ell+1$, take $\ell$ as the last clean boundary window, provided the charge and operator have already stabilized. 
	
	\textit{Symmetry-resolved classification.---}%
	Suppose the commuting onsite generators $U_i$ have orders $n_i$. A sector $\bm q=(q_1,q_2,\ldots)$ means $U_iOU_i^\dagger=e^{2\pi iq_i/n_i}O$, and the projector is
	\begin{equation}
		\Pi_{\bm q}(O)=\prod_i\left[\frac1{n_i}\sum_{s=0}^{n_i-1}e^{-2\pi iq_is/n_i}U_i^sOU_i^{-s}\right].
		\label{eq:charge_projector_general}
	\end{equation}
	For a boundary-supported $O$, Eq.~\eqref{eq:charge_projector_general} is evaluated with the tensor product of the local matrices $u_{i,j}$ only on the support of $O$. Projection precedes minimization and keeps a degenerate lowest-stiffness space resolved by charge.
	
	For $\mathbb Z_N\times\mathbb Z_N$, anchor the first left endpoint with the leftmost microscopic factor of $U_1$. In the clock chain this factor is $X_0$. Dress only the sites to its right:
	\begin{align}
		W_\mu&=\prod_{j=1}^{\ell-1}X_j^{a_j}Z_j^{b_j},\qquad a_j,b_j=0,\ldots,N-1,\nonumber\\
		O_\mu&=X_0W_\mu\nonumber\\
		\Rightarrow O_\mu(\ell=3)&=X_0X_1^{a_1}Z_1^{b_1}X_2^{a_2}Z_2^{b_2}.
		\label{eq:clock_dressing_basis}
	\end{align}
	The first endpoint is neutral under $U_1$, so we project this anchored space into every sector $(0,q)$ with $q=0,\ldots,N-1$ and solve
	\begin{equation}
		D^{(q)}v=\kappa C^{(q)}v
		\label{eq:projected_eig}
	\end{equation}
	for every $q$. Thus $q$ is scanned, not assumed. Let $\kappa_{\min}=\min_q\kappa_{1,q}$ and let $\mathcal Q_{\min}$ contain every charge tied for that minimum within numerical tolerance.

	For each $q\in\mathcal Q_{\min}$, let $V_q$ span the $m_q$-dimensional lowest-stiffness eigenspace and Hilbert--Schmidt orthonormalize $B_{q,i}=\sum_\mu(V_q)_{\mu i}O_\mu^{(q)}$. Search $B_q(c)=\sum_i c_iB_{q,i}$ with $\|c\|_2=1$ (take $c=1$ if $m_q=1$). Writing $\widehat B=B/\sqrt{\Tr(B^\dagger B)/N^\ell}$, define
	\begin{equation}
		\begin{aligned}
			s_q&=\min_{\|c\|_2=1}
			\|\widehat B_q(c)^\dagger\widehat B_q(c)-\mathds1\|_{\rm op},\\
			\mathcal A_\tau&=\{q\in\mathcal Q_{\min}:s_q\leq\tau_U\}.
		\end{aligned}
		\label{eq:qstar}
	\end{equation}
	Test every tied sector. Once a candidate in sector $q$ meets the tolerance, retain its coefficients $c_q^\star$ and stop further starts there. Accept $q_\star$ only if it passes and every other tied sector is certified to have $s_q>\tau_U$. If several sectors pass, report the result as unresolved; if none passes or another sector remains undecided, strengthen the search or enlarge $\ell$. We use $\tau_U=10^{-8}$ in complex double precision and declare the coefficient iteration converged when the phase-aligned change between successive $c_q$ iterates is below $10^{-12}$; search details are given in the SM~\cite{SM}.
	
	With $v_{\rm rec}=V_{q_\star}c_{q_\star}^\star$ for the selected sector, the recovered first left endpoint is
	\begin{equation}
		A_1^\ell=\sum_\mu(v_{\rm rec})_\mu O_\mu^{(q_\star)}.
		\label{eq:Arecover}
	\end{equation}
	Its charge is fixed by construction because every basis element lies in $(0,q_\star)$. With the right-shift convention $X_j\ket{k}=\ket{k+1\bmod N}$ and $Z_jX_j=\omega_NX_jZ_j$, $\omega_N=e^{2\pi i/N}$, the SPT label is
	\begin{equation}
		p=-q_\star\ {\rm mod}\ N.
		\label{eq:pread}
	\end{equation}
	This is the standard projective edge algebra expressed as a directly measured endpoint charge~\cite{Chen2011prb,Schuch2011,Pollmann2012,Else2014}. Reversing the shift convention reverses the sign in Eq.~\eqref{eq:pread}.
	
	The second endpoint operator provides an independent check rather than the primary class readout. Anchor the leftmost microscopic factor $X_1$ of $U_2$ and begin its dressing on site $2$; for the same three-site boundary window,
	\begin{equation}
		O_\nu^{(2)}(\ell=3)=X_1X_2^{a_2}Z_2^{b_2}.
		\label{eq:A2_example}
	\end{equation}
	Project this space into sectors $(r,0)$, recover $A_2^\ell$, and verify on the common boundary window that
	\begin{equation}
		A_1^\ell A_2^\ell\simeq\omega_N^pA_2^\ell A_1^\ell.
		\label{eq:edge_algebra_check}
	\end{equation}
	Thus $q_\star$ and $p$ come from $A_1^\ell$, while $A_2^\ell$ tests the same projective class independently.

	The same symmetry-preserving boundary perturbations in Eq.~\eqref{eq:boundary_perturbations} are applied after classification. A robust endpoint may dress within the boundary window as $h_\partial$ varies, while its selected charge $q_\star$ and projective algebra remain stable. The same perturbation therefore tests both the detected boundary memory and the recovered SPT class without changing the protecting symmetry.

	For the spin-1 $D_2\cong\mathbb Z_2\times\mathbb Z_2$ symmetry, a sector $(q_x,q_z)$ means $U_xOU_x^\dagger=(-1)^{q_x}O$ and $U_zOU_z^\dagger=(-1)^{q_z}O$. After the detection stage has established protected boundary memory, we solve the generalized eigenproblem in all three nontrivial sectors. The sectors $(0,1)$ and $(1,0)$ reconstruct endpoint operators $A_x^\ell$ and $A_z^\ell$, whose algebra defines the spin-1 projective label $p_{D_2}\in\{0,1\}$ through $A_x^\ell A_z^\ell\simeq(-1)^{p_{D_2}}A_z^\ell A_x^\ell$. Thus $p_{D_2}=1$ is the nontrivial Haldane class, while $p_{D_2}=0$ is the trivial projective class. The $(1,1)$ sector reconstructs $A_y^\ell$ and provides an independent redundancy check; one nontrivial sector alone is not sufficient to determine the projective class. The same generalized eigenproblem is evaluated for large-$D$ as a control, but no protected class is assigned when the preceding detection tests fail. The complete spin-1 construction is given in the SM~\cite{SM}.
	
	\textit{Exact cluster benchmark.---}%
	The cluster Hamiltonian in Eq.~\eqref{eq:Hcluster} has a gapped SPT phase for $0\leq\lambda<1$, becomes gapless at the critical point $\lambda=1$, and enters a gapped trivial paramagnet for $\lambda>1$ in this normalization~\cite{Son2011,Lahtinen2015Dec,Verresen2017}. Starting from $O_0=Z_0$, Pauli-string closure gives, before finite-size termination,
	\begin{align}
		b_{2m-1}&=2\lambda,\qquad b_{2m}=2,\nonumber\\
		\Zz_K&=\frac{1-\lambda^2}{1-\lambda^{2K+2}},\qquad \xi_{\rm Krylov}=\frac{1}{|\ln\lambda|}.
		\label{eq:ZKexact}
	\end{align}
	Hence $\Zz_K\to1-\lambda^2$ throughout the SPT phase, while exactly at $\lambda=1$ it becomes $\Zz_K=1/(K+1)$ and vanishes with depth. The SPT-to-trivial bulk transition therefore appears in operator space as a boundary localization-delocalization transition of the Krylov zero mode: its localization length diverges as $\xi_{\rm Krylov}\sim|1-\lambda|^{-1}$, giving $\nu_K=1$, the Ising correlation-length exponent. In this solvable benchmark the topological boundary mode, the bulk gap closing, and Krylov localization are tied by one analytic sequence of Lanczos hoppings. The derivation is given in the SM~\cite{SM}.
	
	\textit{Clock benchmark for the phase label.---}%
	A clock spin is an $N$-level degree of freedom with $X_j\ket{k}=\ket{k+1\bmod N}$, $Z_j\ket{k}=\omega_N^k\ket{k}$, $X_j^N=Z_j^N=\mathds1$, and $Z_jX_j=\omega_NX_jZ_j$. The commuting stabilizers in Eq.~\eqref{eq:clockH} preserve $U_1$ and $U_2$ and realize every $\mathbb Z_N\times\mathbb Z_N$ class $p\in\mathbb Z_N$. Their exact left endpoints are
	\begin{equation}
		A_1=X_0Z_1^p,\qquad A_2=X_1Z_2^{-p},\qquad U_2A_1U_2^\dagger=\omega_N^{-p}A_1.
		\label{eq:fixedpointclass}
	\end{equation}
	These operators and their projective algebra are known analytically from the fixed-point construction. The numerical test is deliberately stricter: for each benchmark Hamiltonian $H_p$, the classifier is supplied the Hamiltonian and local onsite symmetry matrices, scans all charges and boundary dressings, and must recover the hidden label $p$ from Eq.~\eqref{eq:pread}. For $N=4$ it returns $X_0$, $X_0Z_1$, $X_0Z_1^2$, and $X_0Z_1^3$ for $p=0,1,2,3$, respectively, with charges $q_\star=0,3,2,1$. Thus the numerical output reproduces both the exact endpoint operators and the full projective class table from local operator data alone. The table is in the End Matter and the derivation for general $N$ is in the SM~\cite{SM}.
	
	\textit{Conclusion and outlook.---}%
	The two parts of Krylov edge spectroscopy can be summarized as
	\begin{widetext}
		\small
		\begin{equation}
			\begin{aligned}
				\textit{Detection:}\quad &(H,\beta,\{u_{i,j}\})\longrightarrow\{O_0\}\xrightarrow{\Ll\text{-Lanczos}}\{b_n\}\xrightarrow{\eqref{eq:intro_ZK}}\Zz_K\xrightarrow{R_K,B_K,\,V_\partial}\text{protected edge memory},\\
				\textit{Classification:}\quad &(H,\beta,\{u_{i,j}\})\longrightarrow\{O_\mu^{(\bm q)}\}\xrightarrow{\eqref{eq:generalized}}\{\text{slow symmetry sectors}\}\longrightarrow\{\text{recovered endpoints}\}\xrightarrow[\text{edge algebra}]{\text{charge and}}\text{projective class}.
			\end{aligned}
			\label{eq:full_protocol}
		\end{equation}
	\end{widetext}
	
	The Hamiltonian and onsite symmetry matrices generate the local operator search, while $\beta$ selects the energy window only through the inner product in Eq.~\eqref{eq:inner}. The cluster benchmark maps the SPT transition to a Krylov localization-delocalization transition, the clock benchmark reconstructs the cohomology label, and the spin-1 comparison shows why finite $\beta$, geometry, and symmetry-preserving boundary perturbations matter away from fixed points. No ground-state search, entanglement spectrum, or many-body eigenspectrum is required. At fixed Krylov depth, locality gives exact size independence at $\beta=0$ once the commutator support fits inside the system and exponentially small finite-size corrections at finite $\beta$. Together these results provide a reproducible route from local operator dynamics to topological diagnosis and suggest extensions to fermionic, antiunitary, Floquet, and higher-dimensional boundary algebras~\cite{Bultinck2017fMPS,TurzilloYou2019,ElseNayak2016Floquet,Williamson2016MPO}.
	
	\textit{Acknowledgments.---} R.J. thanks Abhinav Prem for fruitful discussions and acknowledges partial support by the U.S. Department of Energy, Office of Science, Office of Advanced Scientific Computing Research via the Exploratory Research for Extreme Scale Science (EXPRESS) program under Award Number DE-SC0026337. H.G.M. was funded by the Deutsche Forschungsgemeinschaft (DFG, German Research Foundation), 436382789, 493420525, and 499180199, via FOR 5522 and large-equipment grants (GOEGrid cluster).
	H. G. M. acknowledges the hospitality of University of Southern California in Los Angeles, USA where a part of this research was performed.
	
	\textit{Data availability---.} The data that support the findings of this article are openly available \cite{this_zenodo}.
	
	\bibliography{refs_kr}
	
	\clearpage
	\onecolumngrid
	\begin{center}
		{\Large{\textbf{End Matter}}}
	\end{center}
	\twocolumngrid
	
\begin{figure}[htbp]
		\centering
		\includegraphics{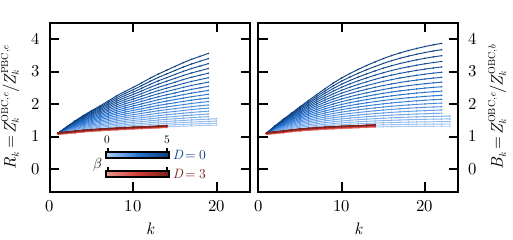}
		\caption{Depth flow of the spin-1 boundary-selectivity ratios at $L=8$ as $\beta$ varies from $0$ to $5$. Blue curves show AKLT ($D=0$) and red curves large-$D$ ($D=3$); increasing $\beta$ enhances the AKLT boundary contrast while the large-$D$ response remains near unity. The left and right panels show $R_K=\Zz_K^{\rm OBC}/\Zz_K^{\rm PBC}$ and $B_K=\Zz_K^{\rm boundary}/\Zz_K^{\rm bulk}$, respectively.}
		\label{fig:detection_kflow}
	\end{figure}

\begin{figure}[htbp]
		\centering
		\includegraphics{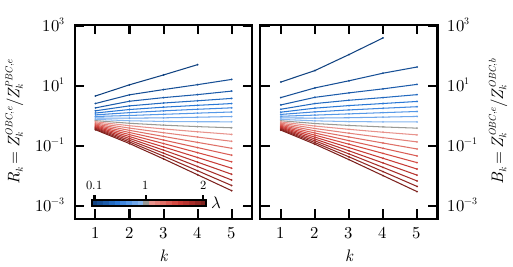}
		\caption{Depth flow of the cluster-chain boundary-selectivity ratios at $L=12$ as $\lambda$ is varied across the transition at $\lambda=1$. In the SPT regime $\lambda<1$, $R_K$ and $B_K$ grow strongly with depth, whereas in the trivial regime $\lambda>1$ they remain near or below unity. The left and right panels show $R_K=\Zz_K^{\rm OBC}/\Zz_K^{\rm PBC}$ and $B_K=\Zz_K^{\rm boundary}/\Zz_K^{\rm bulk}$, respectively.}
		\label{fig:cluster_detection_kflow}
	\end{figure}

	Figures~\ref{fig:detection_kflow} and \ref{fig:cluster_detection_kflow} show the spin-1 and cluster detection flows, respectively. Table~\ref{tab:spin1class} gives the spin-1 classification boundary window flow, and Table~\ref{tab:class} gives the exact $N=4$ clock readout. Classification is interpreted only after the preceding detection tests establish protected boundary memory; otherwise the generalized eigenvalues are retained only as variational stiffnesses. For $p=0$, the anchored clock search returns $A_1=X_0$ already at $\ell=1$.
	
	\newpage	
	
	\begin{table}[H]
		\caption{Boundary window flow of the lowest nontrivial spin-1 branch at $L=8$ and $\beta=5$. Rotational symmetry makes the three AKLT sectors equivalent; the displayed large-$D$ sector is its lowest nontrivial branch. The generalized eigenvalue flow is evaluated for both models, while a projective class is assigned only after the preceding detection stage has established protected boundary memory.}
		\label{tab:spin1class}
		\begin{ruledtabular}
			\begin{tabular}{lccc}
				Model & sector & $\ell$ & $\kappa_1^{\rm OBC}$\\
				\hline
				AKLT & $(1,0)$ & $1$ & $1.12$\\
				AKLT & $(1,0)$ & $2$ & $6.20{\times}10^{-2}$\\
				AKLT & $(1,0)$ & $3$ & $7.19{\times}10^{-3}$\\
				Large-$D$ & $(0,1)$ & $1$ & $7.93$\\
				Large-$D$ & $(0,1)$ & $2$ & $3.13$\\
				Large-$D$ & $(0,1)$ & $3$ & $1.77$
			\end{tabular}
		\end{ruledtabular}
	\end{table}
	
	\begin{table}[H]
		\caption{Exact first-left-endpoint classification for the $\mathbb Z_4\times\mathbb Z_4$ fixed point. The charge $\omega_4^q$ of $A_1$ is measured under $U_2$.}
		\label{tab:class}
		\begin{ruledtabular}
			\begin{tabular}{lccc}
				True $p$ & recovered endpoint & $q_\star$ & $p=-q_\star\ {\rm mod}\ 4$\\
				\hline
				$0$ & $X_0$ & $0$ & $0$\\
				$1$ & $X_0Z_1$ & $3$ & $1$\\
				$2$ & $X_0Z_1^2$ & $2$ & $2$\\
				$3$ & $X_0Z_1^3$ & $1$ & $3$
			\end{tabular}
		\end{ruledtabular}
	\end{table}

\end{document}


\onecolumngrid

\title{Supplemental Material --- Krylov Edge Spectroscopy of Symmetry-Protected Topological Phases}

\author{Heiko Georg Menzler}
\email{heiko.menzler@uni-goettingen.de}
\affiliation{Institute for Theoretical Physics, Georg-August-Universität Göttingen, Friedrich-Hund-Platz 1, 37077 Göttingen, Germany
}

\author{Rishabh Jha}

\email{rishabh.jha@usc.edu}

\affiliation{Department of Physics and Astronomy, University of Southern California, Los Angeles, California 90089-0484, USA}

\maketitle

\tableofcontents

This Supplemental Material supplies the analytical details behind the Letter. We first define the operator inner product and Krylov recursion, prove the exact zero-frequency normalizability criterion and finite-depth leakage bound, and establish the ordered limits in system size and Krylov depth. We then construct a symmetry-complete set of boundary initial operators and symmetry-preserving boundary perturbations, derive the cluster, clock, and balanced Kitaev benchmarks, and develop the symmetry-resolved endpoint classifier. The final section treats degeneracies in the lowest-stiffness space and gives a complete rule for selecting the physical endpoint and its charge.

\section{Operator-space foundations and exact edge-memory criterion}
\label{sec:SM_foundations}

\subsection{Stationary operator Hilbert space}
\label{subsec:SM_metric}

Let $H$ be a finite-range Hamiltonian and choose an inverse temperature $\beta<\infty$. The Gibbs state is
\begin{equation}
\rho_\beta=\frac{e^{-\beta H}}{\Tr(e^{-\beta H})}.
\label{eq:SM_Gibbs}
\end{equation}
It is stationary because $[H,\rho_\beta]=0$. If $U_iHU_i^\dagger=H$ for a protecting symmetry, then $U_i\rho_\beta U_i^\dagger=\rho_\beta$ automatically. Thus $\beta$ fixes the operator-space weighting while $H$ fixes the state. The more general proofs below only require a stationary symmetry-preserving state, so we write $\rho$ when the Gibbs form is unnecessary. Define the connected representative and the symmetrized inner product by
\begin{equation}
\delta A=A-\Tr(\rho A)\mathds 1,
\qquad
(A|B)_S=\frac12\Tr\!\left[\rho\left(\delta A^\dagger\delta B+\delta B\delta A^\dagger\right)\right].
\label{eq:SM_inner}
\end{equation}
The corresponding seminorm is nonnegative. It is a seminorm rather than a norm because $\norm{A}_S$ can vanish for a nonzero operator before null directions are quotiented out:
\begin{equation}
\norm{A}_S^2=\frac12\Tr(\rho\,\delta A^\dagger\delta A)+\frac12\Tr(\rho\,\delta A\delta A^\dagger)\geq0.
\label{eq:SM_norm}
\end{equation}
The operator Hilbert space is the quotient by the null space $\mathcal N_\rho=\{A:\norm{A}_S=0\}$. For full-rank $\rho$, connected subtraction leaves only the identity as a null direction, apart from linear dependencies in a chosen basis. For Hermitian $A$, Eq.~\eqref{eq:SM_norm} is the variance $\Tr(\rho\,\delta A^2)$.

Define the Liouvillian by
\begin{equation}
\Ll A=[H,A].
\label{eq:SM_L}
\end{equation}
Stationarity makes $\Ll$ self-adjoint in Eq.~\eqref{eq:SM_inner}. One of the two trace identities is
\begin{equation}
\Tr\!\left(\rho\,\delta A^\dagger[H,\delta B]\right)
=\Tr\!\left(\rho\,[H,\delta A]^\dagger\delta B\right),
\label{eq:SM_L_trace}
\end{equation}
and the second term in the symmetrized product behaves identically. Therefore
\begin{equation}
(A|\Ll B)_S=(\Ll A|B)_S.
\label{eq:SM_L_self_adjoint}
\end{equation}
Although $\Ll$ maps a Hermitian operator to an anti-Hermitian one, phase choices of the Lanczos vectors absorb this alternation.

Let $O_0=O_0^\dagger$ be connected and normalized. Starting with $O_{-1}=0$ and $b_0=0$, exact Lanczos recursion is
\begin{align}
W_n&=\Ll O_n-b_nO_{n-1},\nonumber\\
a_n&=(O_n|W_n)_S,\nonumber\\
\widetilde W_n&=W_n-a_nO_n,\nonumber\\
b_{n+1}&=\norm{\widetilde W_n}_S,
\qquad
O_{n+1}=\widetilde W_n/b_{n+1}.
\label{eq:SM_Lanczos}
\end{align}
The recursion stops when $b_{n+1}=0$. The generated cyclic space is
\begin{equation}
\mathcal K(O_0)=\overline{\operatorname{span}}\{O_0,\Ll O_0,\Ll^2O_0,\ldots\}.
\label{eq:SM_cyclic}
\end{equation}
All frequency statements below concern only the part of operator space reached from this initial operator.

\subsection{Operator Lanczos algorithm}
\label{subsec:SM_algorithm}

The operator Lanczos algorithm applies the ordinary Lanczos recursion to the linear map $\Ll=[H,\cdot]$ using the inner product in Eq.~\eqref{eq:SM_inner}~\cite{MenzlerJha2024}. It turns repeated commutators into orthonormal operators and a one-dimensional sequence of hopping coefficients. A self-contained implementation is:
\begin{itemize}
\item Choose a Hermitian boundary operator $B$. Form $\delta B=B-\Tr(\rho B)\mathds 1$. If $\norm{\delta B}_S>0$, set $O_0=\delta B/\norm{\delta B}_S$, together with $O_{-1}=0$ and $b_0=0$.
\item At step $n$, compute $W_n=\Ll O_n-b_nO_{n-1}$, then $a_n=(O_n|W_n)_S$, and subtract the component parallel to $O_n$: $\widetilde W_n=W_n-a_nO_n$.
\item Set $b_{n+1}=\norm{\widetilde W_n}_S$. If $b_{n+1}$ vanishes within the chosen arithmetic tolerance, the recursion terminates. Otherwise set $O_{n+1}=\widetilde W_n/b_{n+1}$ and continue.
\item Store $a_n$ and $b_{n+1}$. For the Hermitian initial operators used here, the proof below gives $a_n=0$ in exact arithmetic, so the Krylov representation contains only the off-diagonal hoppings $b_n$.
\end{itemize}
Finite precision can slowly spoil orthogonality. Full reorthogonalization stabilizes the recursion by subtracting every previously constructed direction after each action of $\Ll$:
\begin{equation}
W_n\leftarrow\Ll O_n-\sum_{j=0}^{n}(O_j|\Ll O_n)_S O_j.
\label{eq:SM_full_reorthogonalization}
\end{equation}
Repeating the same subtraction once more further suppresses roundoff when higher accuracy is needed. Then set $b_{n+1}=\norm{W_n}_S$ and $O_{n+1}=W_n/b_{n+1}$. In exact arithmetic, self-adjointness of $\Ll$ makes this full subtraction reduce to the three-term recursion in Eq.~\eqref{eq:SM_Lanczos}; the extra projections only remove roundoff contamination. The recursion therefore requires repeated commutators and inner products. Diagonalizing $H$ is not part of the algorithm. The energy eigenbasis introduced next is used only to interpret the resulting spectral weight.

\subsection{Why a Hermitian initial operator produces a zero-diagonal Krylov chain}
\label{subsec:SM_chiral}

Choose a common eigenbasis $H|m\rangle=E_m|m\rangle$ and $\rho|m\rangle=r_m|m\rangle$. Here and below, a frequency means a Liouvillian eigenvalue $\omega=E_m-E_n$, namely an energy difference probed by the operator. The frequency measure of the normalized initial operator is
\begin{equation}
d\mu_{O_0}(\omega)=\frac12\sum_{m,n}(r_m+r_n)|\langle m|\delta O_0|n\rangle|^2\,\delta\!\left(\omega-(E_m-E_n)\right)d\omega.
\label{eq:SM_measure}
\end{equation}
where $\delta(x)$ is the Dirac delta distribution. Thus $d\mu_{O_0}(\omega)$ records the weight with which $O_0$ connects energy eigenstates separated by $\omega$; it is distinct from the list of energy levels of $H$. Its total weight is one because $O_0$ is normalized. An atom at $\omega=0$ means a nonzero Dirac-delta weight there, equivalently a nondecaying component of the operator autocorrelation. The energy eigenbasis is used here only to define and prove properties of the frequency weight. The protocol itself obtains the same moments and Lanczos coefficients from repeated commutators and the inner product in Eq.~\eqref{eq:SM_inner}.
Hermiticity gives $|\langle m|O_0|n\rangle|^2=|\langle n|O_0|m\rangle|^2$. Exchanging $m$ and $n$ therefore proves
\begin{equation}
d\mu_{O_0}(\omega)=d\mu_{O_0}(-\omega).
\label{eq:SM_even_measure}
\end{equation}
The orthonormal polynomial of degree $n$ associated with an even measure has parity $(-1)^n$. Hence $\omega p_n(\omega)^2$ is odd and
\begin{equation}
a_n=\int\omega p_n(\omega)^2d\mu_{O_0}(\omega)=0.
\label{eq:SM_an_zero}
\end{equation}
The Liouvillian restricted to the Krylov space is consequently a semi-infinite nearest-neighbor hopping matrix with zero diagonal,
\begin{equation}
J=
\begin{pmatrix}
0&b_1&&&\\
b_1&0&b_2&&\\
&b_2&0&b_3&\\
&&b_3&0&\ddots\\
&&&\ddots&\ddots
\end{pmatrix},
\qquad b_n>0
\label{eq:SM_J}
\end{equation}
until a possible exact termination.

Let $J_K$ be the finite matrix obtained from $J$ in Eq.~\eqref{eq:SM_J} by keeping only Krylov sites $0,\ldots,2K$. It contains $2K+1$ sites and therefore has odd dimension. The zero-energy equation $J_K\psi=0$ gives
\begin{equation}
	\psi_{2m+1}=0,
	\qquad
	\psi_{2m}=(-1)^m\alpha_m\psi_0,
	\qquad
	\alpha_m=\prod_{j=1}^{m}\frac{b_{2j-1}}{b_{2j}},
	\qquad
	\alpha_0=1.
	\label{eq:SM_zero_recurrence}
\end{equation}
Normalization defines the finite-depth boundary weight
\begin{equation}
\Zz_K=\left(\sum_{m=0}^{K}|\alpha_m|^2\right)^{-1}.
\label{eq:SM_ZK}
\end{equation}
Every truncation with sites $0,\ldots,2K$ has such a zero eigenvector because the even Krylov sites outnumber the odd Krylov sites by one. The physical question is whether its weight at site zero remains finite as the Krylov chain becomes infinite. As a simple reference case, if every hopping is equal, $b_n=b>0$, then $\alpha_m=1$ for all $m$ and Eq.~\eqref{eq:SM_ZK} gives $\Zz_K=1/(K+1)$, so the finite-chain zero mode delocalizes as $K$ grows.

\subsection{Proof of Theorem 1 in the Main Text}
\label{subsec:SM_theorem_one}

For a nonterminating sequence $b_n>0$, let $J$ initially act on finitely supported sequences in $\ell^2(\mathbb N_0)$, where $\mathbb N_0=\{0,1,2,\ldots\}$ and $\ell^2(\mathbb N_0)$ consists of sequences $\psi=(\psi_0,\psi_1,\ldots)$ satisfying $\sum_{n=0}^{\infty}|\psi_n|^2<\infty$. Assume that the matrix $J$ determines a unique self-adjoint operator on this space, so no additional boundary condition at $n\to\infty$ is required. A simple sufficient, but not necessary, condition for uniqueness is
\begin{equation}
	\sum_{n=1}^{\infty}\frac1{b_n}=\infty.
	\label{eq:SM_Carleman}
\end{equation}
If this series diverges, uniqueness is guaranteed. If it converges, this criterion is inconclusive: uniqueness may still hold or fail.

\begin{theorem}[Krylov normalizability criterion]
\label{thm:SM_normalizability}
Under the assumptions above, the frequency measure of the initial operator has a nonzero atom at zero precisely when the formal zero-energy Krylov wavefunction is square summable:
\begin{equation}
\mu_{O_0}(\{0\})>0
\quad\Longleftrightarrow\quad
\sum_{m=0}^{\infty}\prod_{j=1}^{m}\left(\frac{b_{2j-1}}{b_{2j}}\right)^2<\infty.
\label{eq:SM_normalizability}
\end{equation}
When the series converges,
\begin{equation}
\mu_{O_0}(\{0\})=
\left[\sum_{m=0}^{\infty}\prod_{j=1}^{m}\left(\frac{b_{2j-1}}{b_{2j}}\right)^2\right]^{-1}
=\lim_{K\to\infty}\Zz_K.
\label{eq:SM_atom_value}
\end{equation}
\end{theorem}

\noindent\textit{Proof.} Let $e_n$ denote the standard basis of $\ell^2(\mathbb N_0)$. The projection-valued-measure theorem applied to the self-adjoint realization of the tridiagonal operator $J$ in Eq.~\eqref{eq:SM_J} decomposes it into projectors $E_J(d\omega)$ onto its frequency components. For the cyclic pair $(J,e_0)$, it identifies the previously defined frequency measure with $\langle e_0,E_J(d\omega)e_0\rangle$. Therefore
\begin{equation}
\mu_{O_0}(\{0\})=\norm{P_0e_0}^2,
\qquad
P_0=E_J(\{0\}),
\label{eq:SM_atom_projection}
\end{equation}
so a zero atom exists exactly when $\ker J$ has a vector with nonzero component at $e_0$. A vector $\psi\in\ker J$ satisfies $(J\psi)_n=0$ at every Krylov site. At the boundary site $n=0$, there is no site $-1$, so the first row of $J\psi=0$ contains only $b_1\psi_1=0$. Since $b_1>0$ in the nonterminating case considered here, $\psi_1=0$. At the next even site, $n=2$, the equation is $b_2\psi_1+b_3\psi_3=0$, so $\psi_3=0$. Repeating the same step at $n=4,6,\ldots$ shows inductively that every odd amplitude $\psi_{2m+1}$ vanishes because all $b_n>0$. The remaining equations, at the odd sites $n=2m-1$, are
\begin{equation}
	b_{2m-1}\psi_{2m-2}+b_{2m}\psi_{2m}=0,
	\label{eq:SM_odd_equation}
\end{equation}
so $\psi_{2m}=-(b_{2m-1}/b_{2m})\psi_{2m-2}$. Iterating this relation gives Eq.~\eqref{eq:SM_zero_recurrence}. The even-site equations are then automatically satisfied because they contain only odd amplitudes, all of which vanish. Thus the zero-mode equations split into two simple steps: the boundary condition forces $\psi_1=0$ and hence $\psi_3=\psi_5=\cdots=0$, while the equations on the odd sites recursively determine $\psi_2,\psi_4,\ldots$ from $\psi_0$. This is the origin of the even-site zero-mode profile in Eq.~\eqref{eq:SM_zero_recurrence}.
If $\psi_0=0$, recurrence forces every amplitude to vanish, so $\ker J$ has dimension at most one. The formal solution belongs to $\ell^2$ exactly when the series in Eq.~\eqref{eq:SM_normalizability} converges. Unique self-adjointness identifies this square-summable maximal-domain solution with a vector in the domain of $J$, so it is an actual eigenvector. Its normalization gives $|\psi_0|^2$ equal to the inverse series. Equation~\eqref{eq:SM_atom_projection} then gives Eq.~\eqref{eq:SM_atom_value}. Conversely, a zero atom supplies a nonzero vector in $\ker J$, and the same recurrence forces the displayed series to converge. \hfill$\square$

Equation~\eqref{eq:SM_atom_value} coincides with the stationary correlation value of Ref.~\cite{Gamayun2025Jul} [their Eq.~(5.23)], obtained there by deforming a decaying correlation function and inverting the induced odd-even alternation of the Lanczos coefficients.

If the Lanczos recursion terminates at dimension $D$, use the finite tridiagonal Krylov matrix directly. Positive hoppings give a zero eigenvalue when $D$ is odd and no zero eigenvalue when $D$ is even. For $D=2K+1$, its initial-operator weight is Eq.~\eqref{eq:SM_ZK}. The special case $b_1=0$ has $D=1$ and $O_0$ is already conserved.

The zero-mode coefficient $\alpha_m$ is a product of hopping ratios, so its growth or decay is clearer after taking logarithms. Define
\begin{equation}
	s_m=\ln\frac{b_{2m-1}}{b_{2m}},
	\qquad
	S_m=\sum_{j=1}^{m}s_j,
	\qquad
	S_0=0.
	\label{eq:SM_staggering}
\end{equation}
Then $\alpha_m=e^{S_m}$, so $S_m$ is the cumulative logarithmic change of the zero-mode amplitude and
\begin{equation}
	\frac{|\psi_{2m}|}{|\psi_0|}=e^{S_m},
	\qquad
	\Zz_K=\left(\sum_{m=0}^{K}e^{2S_m}\right)^{-1},
	\qquad
	\mu_{O_0}(\{0\})>0\Longleftrightarrow\sum_{m=0}^{\infty}e^{2S_m}<\infty.
	\label{eq:SM_staggering_criterion}
\end{equation}
This form makes the normalizability test transparent. If $\limsup_{m\to\infty}S_m/m<0$, then $S_m$ eventually decreases at least linearly and $e^{2S_m}$ is bounded by a decaying geometric sequence. If $S_m=-c\ln m+O(1)$, then $e^{2S_m}$ differs from $m^{-2c}$ only by bounded multiplicative factors, so the series converges precisely for $c>1/2$. A negative $s_m$ decreases the zero-mode amplitude across that step, while a positive $s_m$ increases it; normalizability is determined by the accumulated $S_m$ and the full sum in Eq.~\eqref{eq:SM_staggering_criterion}.

\subsection{Memory and near-zero bounds}
\label{subsec:SM_memory_limits}

The normalized connected autocorrelation is the Fourier transform of this frequency measure,
\begin{equation}
C_{O_0}(t)=(O_0|e^{\ii t\Ll}O_0)_S=\int_{\mathbb R}e^{\ii t\omega}d\mu_{O_0}(\omega).
\label{eq:SM_autocorrelation}
\end{equation}
Dominated convergence applied to the time-averaging kernel gives
\begin{equation}
\lim_{T\to\infty}\frac1T\int_0^T C_{O_0}(t)dt=\mu_{O_0}(\{0\}).
\label{eq:SM_long_time}
\end{equation}
Thus Theorem~\ref{thm:SM_normalizability} is exactly a persistent-memory theorem in the Krylov space generated by the initial operator.

For any normalized operator $A$, define its commutator stiffness
\begin{equation}
\kappa(A)=\norm{[H,A]}_S^2=\int\omega^2d\mu_A(\omega).
\label{eq:SM_stiffness}
\end{equation}
For every $\Omega>0$, splitting the second moment into $|\omega|<\Omega$ and $|\omega|\geq\Omega$ gives
\begin{equation}
\mu_A(\{|\omega|\geq\Omega\})\leq\frac{\kappa(A)}{\Omega^2}.
\label{eq:SM_Chebyshev}
\end{equation}
If $A$ is Hermitian, its measure is even and $C_A(t)=\int\cos(\omega t)d\mu_A(\omega)$. Since $\cos x\geq1-x^2/2$,
\begin{equation}
C_A(t)\geq1-\frac{t^2\kappa(A)}2.
\label{eq:SM_time_bound}
\end{equation}
These inequalities are finite-depth statements about near-zero weight. An exact atom requires the infinite-depth condition in Theorem~\ref{thm:SM_normalizability}.

\subsection{Numerical implementation}
\label{subsec:numerical_implementation}

As the numerical implementation of the Lanczos algorithm usually suffers from insurmountable instability problems, we have to carefully convince ourselves that the studied coefficients safely lie in a numerically stable region.
For this we firstly want to guarantee that the basis of the Lanczos operators remains orthogonal throughout the iteration.
The state-of-the-art technique we use is a form of the partial-reorthogonalization technique where Lanczos operators are orthogonalized against the previously calculated operators already collected in the basis only when orthogonality has degraded (see e.g., \cite{JhaMenzler2026lr} for a discussion).
In practice, we do not calculate Lanczos coefficients deep into the iteration and therefore the re-orthogonalization problem is minimized.
However, we still calculate the orthogonality of the basis for every single run and exclude $b_n$ which have been calculated from a degraded basis.

Even having guaranteed that used Lanczos basis vectors are numerically orthogonal to each other, we still can not be sure of the numerical correctness of the Lanczos algorithm.
To improve our confidence in the Lanczos coefficients we run each Lanczos iteration twice, using two formally equal but numerically distinct methodologies.
Comparing the two algorithms allows us to isolate data points which have accumulated too much non-physical noise and exclude them from our analysis.

For these reasons, our presented dataset at times features sequences of Lanczos coefficients which terminate early, because we require that all results have to be derived from trustworthy iteration steps.
Notably, including unstable Lanczos coefficients in the analysis would not qualitatively change any results, indicating that numerical stability degrades before the symmetry arguments are invalidated on which our analysis relies.

We briefly describe the alternative Lanczos algorithm which we use to (in)validate our Lanzcos iteration.
We write the Hamilonian $H$ in its eigenbasis $V$ and use this eigenbasis to diagonalize $\mathcal{L}$.
We write the initial operator as $\tilde{O}_0 = V^\dagger O_0 V$.
Based on this we find 
\begin{align}
    \tilde{O}_{n+1} 
    = \mathcal{L} \tilde{O}_n 
    = [H, \tilde{O}_n] 
    = H V^\dagger O_n V - V^\dagger O_n V H
    = V^\dagger \left(\sum_{\mu, \nu} \omega_{\mu\nu} O_n \right) V\,,
\end{align}
where $\omega_{\mu\nu} = E_\mu - E_\nu$ and $E_k$ being eigenvalues of $H$.
This form of the Lanczos recursion is equivalent to performing Lanczos on the space of polynomials which was already introduced in Section~\ref{subsec:SM_chiral}.

The main difference of this algorithm to the straightforward iteration is that the Lanzcos recursion is performed in the eigenbasis of the Hamiltonian $H$.
The diagonalization incurs a numerical error right at the beginning of the recursion and therefore the algorithm provides a benchmark which is expected to always perform worse than the straightforward recursion.
Thus, when both Lanczos iterations stop agreeing to high precision, we know that we have reached a regime of numerical instability and we discard subsequent results from the iteration.
This check requires the exact diagonalization of the Hamiltonian, which is however not required in the straightforward algorithm.

\section{Ordered convergence in system size and Krylov depth}
\label{sec:SM_convergence}

There are two independent cutoffs. The chain contains $L$ physical sites, while $\Zz_K$ is computed from the Krylov sites $0,\ldots,2K$, a truncation containing $2K+1$ sites. Thus $K$ controls how far the Krylov recursion is followed, whereas $L$ controls the size of the physical chain. The physical order of limits is
\begin{equation}
\Zz_{\rm phys}=\lim_{K\to\infty}\left[\lim_{L\to\infty}\Zz_K(L)\right].
\label{eq:SM_order_limits}
\end{equation}
The inner limit first removes the remote physical boundary while the operator has explored only a fixed neighborhood. The outer limit then asks whether the formal zero mode remains normalizable after arbitrarily many commutators. The next two propositions prove these statements separately.

\subsection{First limit: fixed $K$ and increasing $L$ (Proof of Proposition 1 in the Main Text)}
\label{subsec:SM_fixed_K_L}

Let $\Lambda_L=\{0,\ldots,L-1\}$ and write a finite-range Hamiltonian as
\begin{equation}
	H_L=\sum_{X\subseteq\Lambda_L}h_X,
	\qquad
	h_X=0\ \text{when}\ \operatorname{diam}(X)>R,
	\qquad
	\sup_{j,L}\sum_{X\ni j}\norm{h_X}\leq J.
	\label{eq:SM_finite_range_H}
\end{equation}
We assume size consistency: wherever two finite geometries coincide locally, their interaction terms coincide with those of one fixed infinite-volume interaction. OBC Hamiltonians are restrictions of this interaction and PBC Hamiltonians are its periodizations. Each site has Hilbert-space dimension $d<\infty$. Here $R<\infty$ is the interaction range, $J<\infty$ bounds the interaction strength near one site, and an unsubscripted $\norm{\cdot}$ is the matrix operator norm. The distance $\operatorname{dist}(i,j)$ is the number of nearest-neighbor steps between sites, and $\operatorname{diam}(X)$ is the largest such distance within $X$. For a set $Y$, define its radius-$r$ neighborhood by
\begin{equation}
Y^{[r]}=\{j:\operatorname{dist}(j,Y)\leq r\}.
\label{eq:SM_neighborhood}
\end{equation}
Start from the same local operator $B$ in every system size. For a left-boundary calculation, its support $Y_0$ stays near the left end while the right end recedes. For a bulk or periodic calculation, use coordinates centered on the initial operator so that $Y_0$ is fixed while every physical boundary, or the point where a ring is cut, recedes. At size $L$, connected subtraction and normalization give $O_0^{(L)}=\delta B/\norm{\delta B}_{S,L}$. The limiting geometry is the half-infinite chain for the first case and the infinite chain for the other two.

\begin{convergenceproposition}[Thermodynamic convergence at fixed Krylov depth]
\label{prop:SM_fixed_K_L}
Fix $K<\infty$ and assume that the infinite-volume Lanczos recursion has no breakdown through $b_{2K}$, meaning $b_1(\infty),\ldots,b_{2K}(\infty)>0$. Equivalently, the commutator iterates $B,\Ll B,\ldots,\Ll^{2K}B$ span $2K+1$ independent directions after state-null directions are removed. For the finite-range Hamiltonian in Eq.~\eqref{eq:SM_finite_range_H}:
\begin{align}
\beta=0:&\qquad b_n(L)=b_n(\infty)\quad(1\leq n\leq2K),
\qquad \Zz_K(L)=\Zz_K(\infty),
\label{eq:SM_beta_zero_exact}\\
0<\beta<\infty:&\qquad b_n(L)-b_n(\infty)=O(e^{-\gamma_\beta L})\quad(1\leq n\leq2K),
\qquad \Zz_K(L)-\Zz_K(\infty)=O(e^{-\gamma_\beta L}).
\label{eq:SM_finite_beta_exp}
\end{align}
The first line holds exactly once $Y_0^{[2KR]}$ fits in the chosen geometry without reaching the remote boundary or wrapping around a periodic chain. The second line holds for translation-invariant one-dimensional finite-range Gibbs states with finite-dimensional onsite Hilbert spaces; a finite spatial period is allowed by grouping one period into a larger site. The positive decay rate $\gamma_\beta$ and the prefactors may depend on $K$, $\beta$, the local dimension, and the Hamiltonian, but not on $L$.
\end{convergenceproposition}

The notation $f(L)=O(e^{-\gamma_\beta L})$ has one precise meaning: there are constants $C<\infty$ and $L_0<\infty$, independent of $L$, such that $|f(L)|\leq Ce^{-\gamma_\beta L}$ for every $L\geq L_0$. It does not assert that the same $C$ or $\gamma_\beta$ works after $K$, $\beta$, or the model is changed.

\noindent\textit{Proof.} Define $\operatorname{ad}_{H_L}(A)=[H_L,A]$ and the unnormalized commutator iterates
\begin{equation}
Q_r^{(L)}=\operatorname{ad}_{H_L}^{r}(B),
\qquad
r=0,1,\ldots,2K.
\label{eq:SM_raw_Krylov}
\end{equation}
If an operator $A$ is supported on $Y$, then $[h_X,A]=0$ whenever $X\cap Y=\varnothing$. Every nonzero term in $[H_L,A]$ therefore has support in $Y^{[R]}$. Induction on $r$ gives
\begin{equation}
\operatorname{supp}Q_r^{(L)}\subseteq Y_0^{[rR]}.
\label{eq:SM_support_induction}
\end{equation}
This is an exact algebraic statement. It is a finite-depth commutator cone, rather than a real-time approximation.

All Lanczos data through $b_{2K}$ are obtained by orthogonalizing $Q_0^{(L)},\ldots,Q_{2K}^{(L)}$. Hence they are determined by the finite Gram matrix
\begin{equation}
M_{rs}^{(L)}=(Q_r^{(L)}|Q_s^{(L)})_{S,L},
\qquad
0\leq r,s\leq2K.
\label{eq:SM_raw_Gram}
\end{equation}
Every expectation value in this matrix contains operators supported inside the fixed set $Y_0^{[2KR]}$. At $\beta=0$, $\rho_{0,L}=\mathds 1/d^L$. For any $Y\subseteq\Lambda_L$ (recall $\Lambda_L=\{0,\ldots,L-1\}$) and any operator $A_Y$ supported on $Y$, the normalized trace factorizes exactly over the sites outside $Y$:
\begin{equation}
	\frac1{d^L}\Tr_{\Lambda_L}(A_Y\otimes\mathds 1_{\Lambda_L\setminus Y})
	=\frac1{d^{|Y|}}\Tr_Y(A_Y).
	\label{eq:SM_trace_factorization}
\end{equation}
Thus the infinite-temperature expectation value of a local operator depends only on the sites on which that operator acts, not on how many additional sites are present elsewhere in the chain. Once the commutator cone $Y_0^{[2KR]}$ fits inside the geometry, size consistency also makes the local Hamiltonian terms, and therefore the operators $Q_0^{(L)},\ldots,Q_{2K}^{(L)}$, identical to their thermodynamic-limit counterparts on that region. Their inner products are consequently identical as well, so $M^{(L)}=M^{(\infty)}$ entry by entry. Performing the same finite Lanczos orthogonalization on these identical Gram matrices gives identical $b_1,\ldots,b_{2K}$ and hence identical $\Zz_K$, which proves Eq.~\eqref{eq:SM_beta_zero_exact}.

For $0<\beta<\infty$, set $\rho_{\beta,L}=e^{-\beta H_L}/\Tr(e^{-\beta H_L})$. Unlike at $\beta=0$, the expectation value of a local operator can now depend on Hamiltonian terms elsewhere in the chain through the Gibbs state. In one-dimensional finite-range systems, however, this dependence on a distant boundary is exponentially small~\cite{Araki1969Gibbs,Capel2025Gibbs}. For each fixed finite set $Y$ and each operator $A_Y$ supported there,
\begin{equation}
	\left|\Tr(\rho_{\beta,L}A_Y)-\omega_{\beta,\infty}(A_Y)\right|
	\leq C_{\beta,Y}\norm{A_Y}e^{-\gamma_\beta d(Y,\partial_{\rm far}\Lambda_L)}.
	\label{eq:SM_local_indistinguishability}
\end{equation}
Here $\omega_{\beta,\infty}(A_Y)$ is the thermal expectation value of the same local operator in the corresponding half-infinite or infinite system, and $d(Y,\partial_{\rm far}\Lambda_L)$ is the distance from its support $Y$ to the remote boundary. For fixed $\beta$ and fixed $Y$, the constants $C_{\beta,Y}$ and $\gamma_\beta>0$ are independent of $L$. Thus, as the remote boundary is moved farther from $Y$, the finite-chain expectation approaches its thermodynamic value exponentially in that distance.

For PBC there is no physical boundary, so choose a cut far from $Y$ and remove the finitely many interaction terms crossing that cut. This turns the ring into an open chain while changing the Hamiltonian only far from $A_Y$. Gibbs-state stability makes the effect of this distant change exponentially small~\cite{Capel2025Gibbs}; applying Eq.~\eqref{eq:SM_local_indistinguishability} to the resulting open chain then gives the same exponential estimate for PBC. Applying these bounds to $Q_r$, $Q_r^\dagger Q_s$, $Q_sQ_r^\dagger$, and the one-point functions entering connected subtraction controls every term in the Gram matrix. Since $K$ is fixed, all of these operators have $L$-independent support and norm. Therefore
\begin{equation}
	M_{rs}^{(L)}-M_{rs}^{(\infty)}=O(e^{-\gamma_\beta L})
	\label{eq:SM_Gram_exp}
\end{equation}
for all $0\leq r,s\leq2K$, after reducing $\gamma_\beta$ if necessary.

It remains to pass this estimate through Lanczos orthogonalization. The no-breakdown assumption means that each limiting normalization $b_n(\infty)$ appearing through depth $2K$ is strictly positive. Addition, multiplication, division by a number bounded away from zero, and the positive square root near a positive argument are locally Lipschitz operations: within a sufficiently small neighborhood, an input change of size $\varepsilon$ changes the output by at most a fixed constant times $\varepsilon$. Applying these operations successively to the finitely many entries of $M^{(L)}$ proves by induction that each $b_n(L)-b_n(\infty)$ is $O(e^{-\gamma_\beta L})$. Equation~\eqref{eq:SM_ZK} is a finite expression made from positive $b_1,\ldots,b_{2K}$, so the same argument gives the second estimate in Eq.~\eqref{eq:SM_finite_beta_exp}. If the limiting recursion terminates earlier, use its finite tridiagonal Krylov matrix directly; coefficients beyond the terminating bond are not defined or needed. \hfill$\square$

The estimate in Eq.~\eqref{eq:SM_local_indistinguishability} is a fixed-temperature statement: for every fixed finite $\beta$, the constants controlling the exponential decay may depend on $\beta$ but not on $L$. Increasing $\beta$ gives greater statistical weight to low-energy states and therefore changes the operator metric and, in general, the decay constants, but the same finite-temperature locality argument still applies. The limit $\beta\to\infty$ is different. It replaces the thermal state by a ground-state limit, and when the finite system has degenerate or nearly degenerate ground states, different choices of finite-size ground states can approach different limiting states. A bulk energy gap separates the ground-state sector from excitations but does not by itself select one particular state within that sector or guarantee the boundary-insensitivity estimate needed above. A general zero-temperature convergence statement therefore requires an additional assumption on the chosen ground-state sequence, and is not asserted here.

\subsection{Second limit: increasing $K$ after $L\to\infty$}
\label{subsec:SM_K_limit}

After taking the first limit, write $b_n=\lim_{L\to\infty}b_n(L)$ and define
\begin{equation}
\alpha_0=1,
\qquad
\alpha_m=\prod_{j=1}^{m}\frac{b_{2j-1}}{b_{2j}},
\qquad
S_K=\sum_{m=0}^{K}|\alpha_m|^2,
\qquad
\Zz_K=S_K^{-1}.
\label{eq:SM_K_limit_definitions}
\end{equation}
Increasing $K$ reveals more of the same thermodynamic Lanczos sequence. It does not recompute the earlier $b_n$.

\begin{convergenceproposition}[Krylov-depth convergence]
\label{prop:SM_K_convergence}
For a nonterminating thermodynamic recursion with $b_n>0$, the sequence $\Zz_K$ decreases and has a limit $\Zz_\infty\in[0,1]$. More precisely,
\begin{equation}
\Zz_\infty>0
\quad\Longleftrightarrow\quad
S_\infty\equiv\sum_{m=0}^{\infty}|\alpha_m|^2<\infty,
\qquad
\Zz_\infty=S_\infty^{-1}.
\label{eq:SM_K_limit_iff}
\end{equation}
If constants $A<\infty$ and $0<q<1$ satisfy $|\alpha_m|^2\leq Aq^m$, then
\begin{equation}
0\leq\Zz_K-\Zz_\infty
\leq\frac{Aq^{K+1}}{1-q}
=O(q^K).
\label{eq:SM_K_exp_bound}
\end{equation}
For a terminating recursion, the corresponding finite tridiagonal Krylov matrix gives the exact answer described below Eq.~\eqref{eq:SM_atom_value}: if its dimension $D$ is odd it has a zero eigenvalue, while if $D$ is even it does not. For $D=2K+1$, the zero mode has initial-site weight $\Zz_K$ from Eq.~\eqref{eq:SM_ZK}; the special case $b_1=0$ has $D=1$ and means that $O_0$ is already exactly conserved.
\end{convergenceproposition}

The notation $g(K)=O(q^K)$ means that $|g(K)|\leq Cq^K$ for all $K\geq K_0$, with $C$ and $K_0$ independent of $K$. The bound in Eq.~\eqref{eq:SM_K_exp_bound} displays an admissible constant explicitly.

\noindent\textit{Proof.} Every summand in $S_K$ is nonnegative, so $S_{K+1}\geq S_K\geq1$. Taking reciprocals gives $1\geq\Zz_K\geq\Zz_{K+1}\geq0$. Every bounded monotone real sequence converges, so $\Zz_K\to\Zz_\infty\in[0,1]$. If $S_\infty<\infty$, continuity of $x\mapsto1/x$ on $[1,\infty)$ gives $\Zz_\infty=1/S_\infty>0$. If $S_\infty=\infty$, then $1/S_K\to0$. This proves Eq.~\eqref{eq:SM_K_limit_iff} and reproduces the normalizability criterion in Theorem~\ref{thm:SM_normalizability}.

Under the geometric bound,
\begin{equation}
0\leq S_\infty-S_K=\sum_{m=K+1}^{\infty}|\alpha_m|^2
\leq A\sum_{m=K+1}^{\infty}q^m
=\frac{Aq^{K+1}}{1-q}.
\label{eq:SM_geometric_tail}
\end{equation}
Since $S_K,S_\infty\geq1$,
\begin{equation}
\Zz_K-\Zz_\infty
=\frac{S_\infty-S_K}{S_KS_\infty}
\leq S_\infty-S_K,
\label{eq:SM_reciprocal_tail}
\end{equation}
which proves Eq.~\eqref{eq:SM_K_exp_bound}. \hfill$\square$

Equation~\eqref{eq:SM_K_limit_iff} guarantees convergence in $K$ for every thermodynamic Lanczos sequence. It does not assign a universal rate. A geometric tail gives exponential convergence, while a divergent series may approach zero algebraically or at another rate. No bulk-gap assumption was used, and a bulk gap alone does not impose a geometric tail on the operator Krylov coefficients.

\subsection{Exactly which conclusions apply to the models}
\label{subsec:SM_convergence_models}

All bosonic chains in the Letter have finite local dimension and finite interaction range. The cluster and AKLT families have a one-site translation period. The alternating clock family has period two, which becomes period one after two neighboring qudits are grouped into one larger site. Therefore Convergence proposition~\ref{prop:SM_fixed_K_L} applies as follows:
\begin{equation}
\begin{aligned}
\beta=0:&\quad b_n(L)=b_n(\infty),\quad \Zz_K(L)=\Zz_K(\infty),\\
0<\beta<\infty:&\quad b_n(L)-b_n(\infty)=O(e^{-\gamma_\beta L}),\quad 1\leq n\leq2K,\\
&\quad \Zz_K(L)-\Zz_K(\infty)=O(e^{-\gamma_\beta L}).
\end{aligned}
\label{eq:SM_all_models_L}
\end{equation}
The $\beta=0$ identities become exact after the fixed commutator cone fits, and the finite-temperature estimates hold at every fixed $K$. These statements apply to cluster plus paramagnet, AKLT, large-$D$, and clock chains for every fixed finite $N$ and every $p\in\mathbb Z_N$. At finite temperature, the same fixed-depth conclusion also applies after adding a fixed symmetry-preserving perturbation near the observed boundary. Such a term changes the Hamiltonian only in a region whose size stays fixed as $L$ grows, while the opposite boundary continues to recede. The local Gibbs expectations entering the finite Gram matrix therefore retain exponentially small dependence on that remote boundary. The constants can differ among models, charge sectors, and boundary perturbations. The balanced finite-range Kitaev benchmark obeys an even stronger fixed-depth statement. In the Majorana-linear Krylov sector, commutation with the quadratic Hamiltonian preserves linearity in the Majorana operators, while the symmetrized inner product reduces to the ordinary Euclidean inner product of their coefficient vectors for any stationary parity-preserving state, as shown explicitly in Sec.~\ref{subsec:SM_Kitaev}. Consequently, at fixed Krylov depth the Gram matrix depends only on the finite set of quadratic couplings reached by the commutator cone. Before that cone reaches the opposite boundary, increasing $L$ therefore leaves the corresponding Lanczos coefficients exactly unchanged, independently of temperature. This provides the fermionic counterpart of the locality argument above; related Krylov signatures of boundary and bulk structure in Kitaev chains are developed in Ref.~\cite{JhaMenzler2026lr}.

The stronger $K$-rates are model dependent. For the cluster chain, Eq.~\eqref{eq:SM_cluster_Z} gives
\begin{equation}
\begin{array}{c|c}
0\leq\lambda<1&\Zz_K-(1-\lambda^2)=O(\lambda^{2K})\quad(\lambda=0\text{ terminates exactly})\\
\lambda=1&\Zz_K=(K+1)^{-1}\\
\lambda>1&\Zz_K=O(\lambda^{-2K})
\end{array}
\label{eq:SM_cluster_K_rates}
\end{equation}
for every stationary symmetry-preserving state in which the displayed strings have nonzero norm. For the unperturbed clock fixed point, every nonzero Hermitian quadrature of $Z_0$ commutes with $H_p^{\rm OBC}$ for every finite $N$ and every $p$, so $b_1=0$ and the Krylov chain terminates at dimension one. Thus $\Zz=1$ exactly whenever the initial operator has nonzero connected norm. For perturbed clock, AKLT, and large-$D$, Convergence proposition~\ref{prop:SM_K_convergence} guarantees the monotone $K$ limit at $\beta=0$ and every fixed finite $\beta$. A general exponential rate in $K$, or a positive limiting value for an arbitrary initial operator, requires an additional tail estimate and is not asserted. This is the complete model-by-model scope of the two convergence propositions.

\section{Finite-depth reconstruction, automatic initial operators, and boundary controls}
\label{sec:SM_detection}

\subsection{Proof of Proposition 2 in the Main Text}
\label{subsec:SM_proposition_one}

Theorem~\ref{thm:SM_normalizability} asks whether an infinite sequence is normalizable. The following finite-depth construction answers a different question: how much the sharply truncated candidate leaks into the next Krylov site.

\begin{proposition}[Finite-depth Krylov leakage]
\label{prop:SM_leakage}
Let $O_0,\ldots,O_{2K}$ exist. Define
\begin{equation}
A_K=\sqrt{\Zz_K}\sum_{m=0}^{K}(-1)^m\alpha_mO_{2m},
\qquad
\alpha_0=1,
\qquad
\alpha_m=\prod_{j=1}^{m}\frac{b_{2j-1}}{b_{2j}}.
\label{eq:SM_AK}
\end{equation}
Then $A_K=A_K^\dagger$ and $\norm{A_K}_S=1$. One further Lanczos step gives
\begin{align}
\Ll A_K&=(-1)^Kb_{2K+1}\alpha_K\sqrt{\Zz_K}\,O_{2K+1},\nonumber\\
\varepsilon_K&\equiv\norm{[H,A_K]}_S=b_{2K+1}|\alpha_K|\sqrt{\Zz_K},\nonumber\\
\kappa(A_K)&=\varepsilon_K^2,
\label{eq:SM_leakage}
\end{align}
when $b_{2K+1}>0$. If the recursion terminates at this step, $b_{2K+1}=0$, the same cancellation gives $\Ll A_K=0$ and $\varepsilon_K=0$, so $A_K$ is an exact zero-frequency operator.
\end{proposition}

\noindent\textit{Proof.} Starting from Hermitian $O_0$, the even Lanczos vectors can be chosen Hermitian and the odd vectors anti-Hermitian; the coefficients of the even vectors in Eq.~\eqref{eq:SM_AK} are real, so $A_K$ is Hermitian. Orthonormality and Eq.~\eqref{eq:SM_ZK} give $\norm{A_K}_S^2=\Zz_K\sum_{m=0}^{K}|\alpha_m|^2=1$. Since $a_n=0$,
\begin{equation}
\Ll O_n=b_nO_{n-1}+b_{n+1}O_{n+1}.
\label{eq:SM_chiral_recursion}
\end{equation}
For $0\leq m<K$, the coefficient of $O_{2m+1}$ in $\Ll A_K$ is proportional to
\begin{equation}
b_{2m+1}\alpha_m-b_{2m+2}\alpha_{m+1}=0,
\label{eq:SM_telescoping}
\end{equation}
because $\alpha_{m+1}=\alpha_m b_{2m+1}/b_{2m+2}$. Every interior contribution cancels. The sole remainder couples $O_{2K}$ to $O_{2K+1}$ and gives Eq.~\eqref{eq:SM_leakage}. \hfill$\square$

Constructing $A_K$ uses vectors through $O_{2K}$ and hoppings through $b_{2K}$. The coefficient $b_{2K+1}$ is one additional action of $\Ll$ that tests the already-defined operator. Equations~\eqref{eq:SM_Chebyshev} and \eqref{eq:SM_time_bound} applied to $A_K$ give
\begin{equation}
\mu_{A_K}(|\omega|\geq\Omega)\leq\frac{\varepsilon_K^2}{\Omega^2},
\qquad
C_{A_K}(t)\geq1-\frac{t^2\varepsilon_K^2}{2}.
\label{eq:SM_AK_bounds}
\end{equation}
Thus $\varepsilon_K$ certifies this particular finite-depth reconstruction. It is logically independent of the necessary-and-sufficient infinite-depth atom criterion.

\subsection{Automatic symmetry-complete initial-operator set}
\label{subsec:SM_seed_bank}

Consider a qudit with one-site Hilbert space $\mathbb C^d$. Define generalized Weyl matrices by
\begin{equation}
X_d|k\rangle=|k+1\bmod d\rangle,
\qquad
Z_d|k\rangle=\omega_d^k|k\rangle,
\qquad
\omega_d=e^{2\pi\ii/d}.
\label{eq:SM_Weyl}
\end{equation}
The $d^2$ matrices $W_{ab}=X_d^aZ_d^b$, with $a,b=0,\ldots,d-1$, span every one-site operator. Remove $W_{00}=\mathds 1$ and form Hermitian candidates
\begin{equation}
H_{ab}^{(+)}=\frac{W_{ab}+W_{ab}^\dagger}{\sqrt2},
\qquad
H_{ab}^{(-)}=\frac{W_{ab}-W_{ab}^\dagger}{\ii\sqrt2}.
\label{eq:SM_Hermitian_Weyl}
\end{equation}
Zero matrices and linear dependencies are removed with the local Hilbert--Schmidt Gram matrix. The result has exactly $d^2-1$ independent traceless Hermitian directions.

Let a finite abelian symmetry group $G$ act onsite as $U_g=\bigotimes_j u_{g,j}$. A character $\chi$ is a one-dimensional representation satisfying $\chi(gh)=\chi(g)\chi(h)$. The component of a local operator in sector $\chi$ is
\begin{equation}
\Pi_\chi(B)=\frac1{|G|}\sum_{g\in G}\chi(g)^*u_{g,0}Bu_{g,0}^\dagger.
\label{eq:SM_local_charge_projector}
\end{equation}
Character orthogonality gives $\Pi_\chi\Pi_{\chi'}=\delta_{\chi\chi'}\Pi_\chi$. For a complex character, $\Pi_\chi(B)^\dagger$ belongs to the conjugate sector, so the two Hermitian quadratures combine the pair $\chi\oplus\chi^*$. Collect every nonzero Hermitian result and remove dependencies. Retaining every symmetry sector gives a complete one-site Hermitian operator set organized by charge.

For candidate matrices $Q_1,\ldots,Q_M$, define
\begin{equation}
G_{ij}^{\rm loc}=\frac1d\Tr(Q_iQ_j).
\label{eq:SM_local_Gram}
\end{equation}
Diagonalizing this small matrix and retaining its positive eigenspace produces an orthonormal local basis. After embedding at site zero, construct the Gram matrix again in Eq.~\eqref{eq:SM_inner}; quotient any directions invisible in the chosen state and normalize every survivor before Lanczos.

\begin{lemma}[Completeness of the boundary initial-operator set]
\label{lem:SM_seed_completeness}
Let $\mathcal S_0$ be the one-site operator subspace at the left boundary after quotienting state-null directions, let $\{B_a\}$ be any complete orthonormal basis of $\mathcal S_0$, and let $P_0$ project onto $\ker\Ll$ in the operator Hilbert space. If an exact edge operator $A\in\ker\Ll$ has nonzero one-site projection $P_{\mathcal S_0}A$, then at least one initial operator $B_a$ has a nonzero zero-frequency atom.
\end{lemma}

\noindent\textit{Proof.} Completeness gives $P_{\mathcal S_0}A=\sum_aB_a(B_a|A)_S$. If this projection is nonzero, at least one overlap $(B_a|A)_S$ is nonzero. Since $A=P_0A$ and $P_0$ is self-adjoint,
\begin{equation}
(B_a|A)_S=(P_0B_a|A)_S.
\label{eq:SM_seed_overlap}
\end{equation}
Hence $P_0B_a\neq0$, and the zero-frequency atom of the normalized initial operator is $\norm{P_0B_a}_S^2>0$. \hfill$\square$

This is the precise guarantee summarized in the Letter. It is conditional only on one-site visibility. Its contrapositive is also useful: if every element of a complete one-site set has zero projection onto $\ker\Ll$, then no exact zero-frequency edge operator has a nonzero one-site component in that operator space. This does not by itself imply that the phase is trivial; it only shows that an exact edge memory is not visible within the tested one-site space. If the physical edge operator first becomes visible on several sites, enlarge the initial-operator set to all Hermitian Weyl strings on the first $r$ sites and repeat where the enrlargement should not to be to the point that the boundary starts mixing with the bulk. The same proof applies with $\mathcal S_0$ replaced by the $r$-site boundary space. A reduced physically motivated set can be useful, while the completeness guarantee belongs to the full set.

\subsection{Explicit one-site initial-operator sets for the models in the Letter}
\label{subsec:SM_seed_examples}

For the cluster chain, $d=2$ and the one-site matrices are
\begin{equation}
X=\begin{pmatrix}0&1\\1&0\end{pmatrix},
\qquad
Y=\begin{pmatrix}0&-\ii\\\ii&0\end{pmatrix},
\qquad
Z=\begin{pmatrix}1&0\\0&-1\end{pmatrix}.
\label{eq:SM_Pauli}
\end{equation}
The symmetry generators are $U_{\rm e}=\prod_{j\ {\rm even}}X_j$ and $U_{\rm o}=\prod_{j\ {\rm odd}}X_j$. At the even site zero, their local matrices are $(X,\mathds 1)$. Conjugation gives
\begin{equation}
X:(0,0),
\qquad
Y:(1,0),
\qquad
Z:(1,0),
\label{eq:SM_cluster_sectors}
\end{equation}
where charge one under a $\mathbb Z_2$ generator means a minus sign. The automatic one-site set is $\{X_0,Y_0,Z_0\}$. The exact cluster calculation uses the single initial operator $Z_0$.

For the alternating clock chain, $d=N$ and the matrices are Eq.~\eqref{eq:SM_Weyl}. For $N=3$,
\begin{equation}
X_3=\begin{pmatrix}0&0&1\\1&0&0\\0&1&0\end{pmatrix},
\qquad
Z_3=\begin{pmatrix}1&0&0\\0&\omega_3&0\\0&0&\omega_3^2\end{pmatrix}.
\label{eq:SM_clock_matrices}
\end{equation}
At site zero, the two alternating generators act locally as $(X_N,\mathds 1)$. Since
\begin{equation}
X_N(X_N^aZ_N^b)X_N^\dagger=\omega_N^{-b}X_N^aZ_N^b,
\label{eq:SM_clock_charge_Weyl}
\end{equation}
the charge is $(-b\bmod N,0)$. Processing all Weyl matrices gives $N^2-1$ Hermitian initial operators. The two quadratures of $Z_0$ used as a minimal probe in the Letter are
\begin{equation}
\frac{Z_0+Z_0^\dagger}{\sqrt2},
\qquad
\frac{Z_0-Z_0^\dagger}{\ii\sqrt2}.
\label{eq:SM_clock_minimal_seeds}
\end{equation}
For an unbiased scan over an unknown Hamiltonian or class, use the complete set rather than only this pair.

For AKLT and large-$D$, each site is a spin-one qudit with $d=3$. In the basis $\{|1\rangle,|0\rangle,|-1\rangle\}$,
\begin{equation}
S^x=\frac1{\sqrt2}\begin{pmatrix}0&1&0\\1&0&1\\0&1&0\end{pmatrix},
\quad
S^y=\frac1{\sqrt2}\begin{pmatrix}0&-\ii&0\\\ii&0&-\ii\\0&\ii&0\end{pmatrix},
\quad
S^z=\begin{pmatrix}1&0&0\\0&0&0\\0&0&-1\end{pmatrix}.
\label{eq:SM_spin_one_matrices}
\end{equation}
The protecting group $D_2\cong\mathbb Z_2\times\mathbb Z_2$ is generated locally by
\begin{equation}
u_x=e^{\ii\pi S^x}=\begin{pmatrix}0&0&-1\\0&-1&0\\-1&0&0\end{pmatrix},
\qquad
u_z=e^{\ii\pi S^z}=\begin{pmatrix}-1&0&0\\0&1&0\\0&0&-1\end{pmatrix}.
\label{eq:SM_D2_matrices}
\end{equation}
Here $D_2$ denotes the four-element group $\{\mathds 1,U_x,U_z,U_xU_z\}$ of global half-turns. A complete traceless Hermitian basis and its sectors are
\begin{equation}
\begin{array}{c|l}
(q_x,q_z)&\text{one-site operators}\\
\hline
(0,1)&S^x,\ S^yS^z+S^zS^y\\
(1,1)&S^y,\ S^xS^z+S^zS^x\\
(1,0)&S^z,\ S^xS^y+S^yS^x\\
(0,0)&(S^x)^2-(S^y)^2,\ 3(S^z)^2-2\mathds 1
\end{array}
\label{eq:SM_spin_one_bank}
\end{equation}
The full automatic set has eight directions. The minimal set $\{S_0^x,S_0^y,S_0^z\}$ used in the Letter samples all three nontrivial $D_2$ sectors; the complete set also includes the quadrupolar direction in each sector.

\subsection{Geometry controls and symmetry-preserving boundary perturbations}
\label{subsec:SM_detection_controls}

For the same $\beta$, the same normalized initial operator $O_0$, and the same Krylov depth $K$ in each geometry, define
\begin{equation}
R_K=\frac{\Zz_K^{\rm OBC}}{\Zz_K^{\rm PBC}},
\qquad
B_K=\frac{\Zz_K^{\rm boundary}}{\Zz_K^{\rm bulk}}.
\label{eq:SM_RB}
\end{equation}
The absolute open-chain weight and its depth and size flow accompany these ratios. The ratios isolate an open physical boundary from a periodic or bulk response. A symmetry-preserving boundary perturbation then tests whether the signal arises from an accidentally omitted boundary term.

Pins can be generated without model-specific guessing. Choose a boundary range $r$, generate all Hermitian Weyl strings on those sites, and project each trial operator $B$ into the invariant sector,
\begin{equation}
\Pi_{\rm inv}^{\partial}(B)=\frac1{|G|}\sum_{g\in G}u_g^{\partial}B\,u_g^{\partial\dagger},
\qquad
u_g^{\partial}=\bigotimes_{j\in\partial}u_{g,j}.
\label{eq:SM_pin_projector}
\end{equation}
Remove zero directions, the identity, and dependencies with a Hilbert--Schmidt Gram matrix. Every remaining Hermitian direction is an allowed symmetry-preserving boundary perturbation. If this invariant space contains only the identity, increase $r$.

The simplest choices used in the Letter are
\begin{align}
V_{\partial}^{\rm cl}&=X_0+X_{L-1},\nonumber\\
V_{\partial}^{\rm clock}&=\frac12\left(X_0+X_0^\dagger+X_{L-1}+X_{L-1}^\dagger\right),\nonumber\\
V_{\partial}^{D_2}&=\sum_{e\in\{0,L-1\}}\left[(S_e^z)^2-\frac23\mathds 1\right].
\label{eq:SM_simple_pins}
\end{align}
Direct conjugation by the local matrices in Eqs.~\eqref{eq:SM_Pauli}, \eqref{eq:SM_Weyl}, and \eqref{eq:SM_D2_matrices} proves that these operators preserve the stated symmetries.

\begin{lemma}[Symmetric action on an isolated projective edge]
\label{lem:SM_Schur}
Suppose the isolated left-edge Hilbert space carries an irreducible projective representation $V_g$. The projection of a local symmetry-preserving boundary perturbation into this edge space is proportional to the identity.
\end{lemma}

\noindent\textit{Proof.} Symmetry preservation makes the projected perturbation commute with every $V_g$. Projective phases cancel in conjugation. Irreducibility then implies that every operator commuting with all $V_g$ is proportional to the identity. \hfill$\square$

Repeated irreducible blocks may have a splittable multiplicity space, and a finite chain may have exponentially weak coupling between its two edges. The protected projective factor remains. The edge operator may also rotate within the predetermined initial-operator set, which is why the complete set is rerun after the boundary perturbation.

\section{Exact analytic benchmarks and critical scaling}
\label{sec:SM_exact_benchmarks}

\subsection{Cluster chain in a paramagnetic field}
\label{subsec:SM_cluster}

Let
\begin{equation}
H_{\rm cl}(\lambda)=-\sum_{j=1}^{L-2}Z_{j-1}X_jZ_{j+1}-\lambda\sum_{j=0}^{L-1}X_j,
\label{eq:SM_cluster_H}
\end{equation}
with $\lambda\geq0$. It preserves $U_{\rm e}$ and $U_{\rm o}$ defined above. In this normalization, $0\leq\lambda<1$ is the gapped SPT phase, $\lambda=1$ is the gapless SPT-to-trivial critical point, and $\lambda>1$ is the gapped trivial paramagnet~\cite{Son2011,Lahtinen2015Dec,Verresen2017}. Starting from $A_0=Z_0$, introduce
\begin{equation}
A_m=\left(\prod_{r=0}^{2m-1}X_r\right)Z_{2m},
\qquad
B_m=\left(\prod_{r=0}^{2m-1}X_r\right)Y_{2m},
\qquad
B_{-1}=0.
\label{eq:SM_cluster_strings}
\end{equation}
Every string is odd under $U_{\rm e}$, so its expectation vanishes in a symmetry-preserving state. Distinct strings anticommute at the earlier endpoint, and each string squares to the identity. They are therefore orthonormal in Eq.~\eqref{eq:SM_inner} for every stationary state preserving $U_{\rm e}$.

Only Hamiltonian terms overlapping the endpoint of a string can change $A_m$ or $B_m$. Using $XZ=-ZX=-\ii Y$, $XY=-YX=\ii Z$, and their cyclic permutations gives the closed commutator equations
\begin{equation}
\ii\Ll A_m=2B_{m-1}-2\lambda B_m,
\qquad
\ii\Ll B_m=2\lambda A_m-2A_{m+1}.
\label{eq:SM_cluster_closure}
\end{equation}
For $m=0$, $A_0=Z_0$ and the only contributing term is $-\lambda X_0$, so $\ii[H,A_0]=-2\lambda Y_0=-2\lambda B_0$, consistent with $B_{-1}=0$. For general $m$, the field at site $2m$ produces the term proportional to $B_m$, while the neighboring cluster term removes the last two $X$ factors and produces $B_{m-1}$; the second equation follows from the same two local Pauli products. Thus repeated commutation stays exactly inside the displayed string sequence.
A phase gauge makes the Lanczos hoppings positive. Before a remote boundary is reached,
\begin{equation}
b_{2m-1}=2\lambda,
\qquad
b_{2m}=2,
\qquad
m\geq1.
\label{eq:SM_cluster_b}
\end{equation}
Hence $\alpha_m=\lambda^m$ and
\begin{equation}
\Zz_K=
\begin{cases}
\displaystyle\frac{1-\lambda^2}{1-\lambda^{2K+2}},&\lambda\neq1,\\[7pt]
\displaystyle\frac1{K+1},&\lambda=1.
\end{cases}
\label{eq:SM_cluster_Z}
\end{equation}
For $0\leq\lambda<1$, the normalized half-line edge operator is
\begin{equation}
	\Psi_{\rm L}^{(z)}=\sqrt{1-\lambda^2}\sum_{m=0}^{\infty}\lambda^mA_m.
	\label{eq:SM_cluster_edge}
\end{equation}
To verify that it is conserved, apply the first relation in Eq.~\eqref{eq:SM_cluster_closure} to every $A_m$. Apart from the common normalization factor,
\begin{equation}
	\ii[H_{\rm cl},\Psi_{\rm L}^{(z)}]
	=2\sum_{m=0}^{\infty}\lambda^mB_{m-1}
	-2\sum_{m=0}^{\infty}\lambda^{m+1}B_m,
	\qquad B_{-1}=0.
\end{equation}
Shifting the index in the first sum gives $2\sum_{m=0}^{\infty}\lambda^{m+1}B_m$, which cancels the second sum term by term. Hence $[H_{\rm cl},\Psi_{\rm L}^{(z)}]=0$ on the half-line. The coefficient of the initial operator $A_0=Z_0$ is $\sqrt{1-\lambda^2}$, so its squared weight is $\Zz_\infty=1-\lambda^2$. Moreover, the coefficients decay as $\lambda^m=e^{-m|\ln\lambda|}$, giving the Krylov localization length
\begin{equation}
	\Zz_\infty=1-\lambda^2,
	\qquad
	\xi_{\rm K}=\frac1{|\ln\lambda|}.
	\label{eq:SM_cluster_weight_length}
\end{equation}
For $\lambda<1$, the geometric sequence is square summable and the edge operator is normalizable. At $\lambda>1$, the same formal left-edge sequence grows with $m$, so its squared coefficients cannot be summed and no normalizable half-line zero mode of this form exists. At $\lambda=1$, the coefficients neither grow nor decay: $\alpha_m=1$, and Eq.~\eqref{eq:SM_cluster_Z} gives the finite-depth weight exactly as $\Zz_K=(K+1)^{-1}$, which vanishes as $K\to\infty$.

Let $\delta=1-\lambda\to0^+$ as $\lambda\to1^-$ from the topological side. Since $-\ln(1-\delta)=\delta+O(\delta^2)$,
\begin{equation}
\xi_{\rm K}\sim\delta^{-\nu_{\rm K}},
\qquad
\nu_{\rm K}=1,
\qquad
\Zz_\infty\sim2\delta,
\qquad
\sqrt{\Zz_\infty}\sim\sqrt2\,\delta^{1/2}.
\label{eq:SM_cluster_exponents}
\end{equation}
Thus the Krylov localization-length exponent is one, matching the Ising correlation-length exponent of the dual cluster-field problem~\cite{Son2011,Lahtinen2015Dec}. The exponent of $\Zz_\infty$ is one, while the initial-operator-to-edge overlap amplitude has exponent one half. These last two describe boundary zero-frequency weight rather than the conventional bulk magnetization.

\subsection{Balanced nearest-neighbor Kitaev chain}
\label{subsec:SM_Kitaev}

The balanced quadratic Kitaev chain provides a second exact test of the detection theorem. In fermionic variables, take the open-chain Hamiltonian
\begin{equation}
	H_{\rm K}
	=-\mu\sum_{j=0}^{L-1}\left(c_j^\dagger c_j-\frac12\right)
	-t\sum_{j=0}^{L-2}\left(c_j^\dagger c_{j+1}+c_{j+1}^\dagger c_j\right)
	+\Delta\sum_{j=0}^{L-2}\left(c_jc_{j+1}+c_{j+1}^\dagger c_j^\dagger\right),
	\label{eq:SM_Kitaev_fermion_H}
\end{equation}
where $\mu$ is the chemical potential, $t$ is the nearest-neighbor hopping amplitude, and $\Delta$ is the $p$-wave pairing amplitude. After choosing a gauge in which $t$ and $\Delta$ are real, the balanced point is $\Delta=t$. Define Majoranas
\begin{equation}
	\gamma_{A,j}=c_j+c_j^\dagger,
	\qquad
	\gamma_{B,j}=-\ii(c_j-c_j^\dagger),
	\qquad
	\{\gamma_{\alpha,j},\gamma_{\beta,k}\}=2\delta_{\alpha\beta}\delta_{jk}.
\end{equation}
At $\Delta=t$, Eq.~\eqref{eq:SM_Kitaev_fermion_H} becomes
\begin{equation}
	H_{\rm K}^{\rm bal}=-\frac{\ii\mu}{2}\sum_{j=0}^{L-1}\gamma_{A,j}\gamma_{B,j}+\ii t\sum_{j=0}^{L-2}\gamma_{B,j}\gamma_{A,j+1}.
	\label{eq:SM_Kitaev_H}
\end{equation}
For $O(v)=\sum_{j,\alpha}v_{\alpha,j}\gamma_{\alpha,j}$, any stationary parity-preserving state gives
\begin{equation}
(O(v)|O(w))_S=v^\dagger w.
\label{eq:SM_Kitaev_metric}
\end{equation}
Thus the Majorana-linear metric is state independent. Their commutator algebra closes:
\begin{align}
\Ll\gamma_{A,j}&=\ii\mu\gamma_{B,j}+2\ii t\gamma_{B,j-1},\nonumber\\
\Ll\gamma_{B,j}&=-\ii\mu\gamma_{A,j}-2\ii t\gamma_{A,j+1}.
\label{eq:SM_Kitaev_commutators}
\end{align}
Starting from $\gamma_{A,0}$, Lanczos visits $\gamma_{A,0},\gamma_{B,0},\gamma_{A,1},\gamma_{B,1},\ldots$ up to phases, giving
\begin{equation}
b_{2m-1}=|\mu|,
\qquad
b_{2m}=2|t|.
\label{eq:SM_Kitaev_b}
\end{equation}
With $r=|\mu|/(2|t|)$,
\begin{equation}
\Zz_K=
\begin{cases}
\displaystyle\frac{1-r^2}{1-r^{2K+2}},&r\neq1,\\[7pt]
\displaystyle\frac1{K+1},&r=1,
\end{cases}
\qquad
\Gamma_{\rm L}=\sqrt{1-r^2}\sum_{j=0}^{\infty}\left(-\frac{\mu}{2t}\right)^j\gamma_{A,j}.
\label{eq:SM_Kitaev_exact}
\end{equation}
The edge is normalizable for $r<1$, with $\Zz_\infty=1-r^2$ and $\xi_{\rm K}=1/|\ln r|$. The same exponents as Eq.~\eqref{eq:SM_cluster_exponents} follow as $r\to1^-$ from the normalizable side. This exact reduction relies on quadratic closure. With interactions, the many-body theorem and Proposition~\ref{prop:SM_leakage} remain valid, while Eq.~\eqref{eq:SM_Kitaev_b} generally does not~\cite{JhaMenzler2026lr}.

For a nonconstant hopping sequence, the rigorous criterion remains the one derived in Eq.~\eqref{eq:SM_staggering_criterion},
\begin{equation}
	\sum_{m=0}^{\infty}e^{2S_m}<\infty,
	\qquad
	S_m=\sum_{j=1}^{m}s_j,
	\qquad
	s_j=\ln\frac{b_{2j-1}}{b_{2j}}.
\end{equation}
As explained below Eq.~\eqref{eq:SM_staggering}, the zero-mode amplitudes satisfy $|\psi_{2m}|/|\psi_0|=e^{S_m}$. Thus $s_m$ describes the local change of the amplitude from one step to the next, while $S_m$ records the accumulated growth or decay. A sign change of $s_m$, equivalently a crossing of $s_m$ through zero, therefore marks only a local switch between contraction and expansion of the zero-mode profile; it does not by itself determine whether the full profile is square summable. For example, constant $s_m=-a<0$ gives $S_m=-am$ and hence exponential decay $e^{S_m}=e^{-am}$, producing a normalizable mode even though $s_m$ never crosses zero. Conversely, $s_m=(-1)^ma$ changes sign at every step, but its partial sums $S_m$ remain bounded, so $e^{2S_m}$ does not decay and the series diverges. Crossing counts are therefore model-specific compressions of the full integrated-staggering data $S_m$, while Eq.~\eqref{eq:SM_staggering_criterion} is the general normalizability test.

\section{Symmetry-resolved variational classification}
\label{sec:SM_classification}

\subsection{Minimizing the commutator in symmetry sectors}
\label{subsec:SM_variational}

Choose a boundary window of $\ell$ sites and a linearly complete operator set $\{O_a\}$ in the intended search space. Define
\begin{equation}
C_{ab}=(O_a|O_b)_S,
\qquad
D_{ab}=([H,O_a]|[H,O_b])_S.
\label{eq:SM_CD}
\end{equation}
For $O(v)=\sum_av_aO_a$, define the normalized commutator stiffness
\begin{equation}
	\kappa(v)\equiv
	\frac{\norm{[H,O(v)]}_S^2}{\norm{O(v)}_S^2}
	=\frac{v^\dagger Dv}{v^\dagger Cv}.
	\label{eq:SM_Rayleigh}
\end{equation}
For Hermitian $O(v)$, this equals the square of the first Lanczos hopping, $\kappa(v)=b_1^2(v)$\footnote{The general relation is $b_1^2 = \kappa - |a_0|^2$.}. For a general charged operator, the classification uses $\kappa(v)$ directly. The classification step therefore asks for the boundary operator with the smallest normalized commutator stiffness inside each symmetry sector.
Both matrices are positive semidefinite. Diagonalize $C=W\Lambda W^\dagger$, remove its zero-eigenvalue directions, and set $X=W_+\Lambda_+^{-1/2}$. Writing $v=Xy$ gives $v^\dagger Cv=y^\dagger y$, so Eq.~\eqref{eq:SM_Rayleigh} becomes
\begin{equation}
\frac{y^\dagger X^\dagger DXy}{y^\dagger y}.
\label{eq:SM_whitened_ratio}
\end{equation}
Minimize this expression subject to $y^\dagger y=1$. Introducing a multiplier $\kappa$ and differentiating $y^\dagger X^\dagger DXy-\kappa(y^\dagger y-1)$ with respect to $y^\dagger$ gives
\begin{equation}
X^\dagger DXy=\kappa y,
\qquad
v=Xy.
\label{eq:SM_whitened_problem}
\end{equation}
Multiplying back by the retained $C$ basis gives the equivalent generalized eigenproblem
\begin{equation}
Dv=\kappa Cv.
\label{eq:SM_generalized_problem}
\end{equation}
The smallest eigenvalue is therefore the smallest normalized commutator stiffness in the chosen operator space. If $\mathcal S_\ell\subseteq\mathcal S_{\ell+1}$, every trial operator available at window $\ell$ is still available at $\ell+1$. Minimizing over the larger set can only preserve or lower the minimum:
\begin{equation}
\kappa_1(\ell+1)\leq\kappa_1(\ell).
\label{eq:SM_kappa_monotone}
\end{equation}

Let $G$ be a finite abelian onsite symmetry group. An operator has charge $\chi$ when
\begin{equation}
U_gOU_g^\dagger=\chi(g)O.
\label{eq:SM_charge_definition}
\end{equation}
Project before minimization:
\begin{equation}
\Pi_\chi(O)=\frac1{|G|}\sum_{g\in G}\chi(g)^*U_gOU_g^\dagger.
\label{eq:SM_global_projector}
\end{equation}
If $O$ is supported on a boundary window $\partial$, only the local tensor product $u_g^\partial=\bigotimes_{j\in\partial}u_{g,j}$ is needed because the symmetry factors outside $\partial$ commute through $O$ and cancel. Thus the global $U_g$ in Eq.~\eqref{eq:SM_global_projector} is conceptual notation for an operation implemented entirely from onsite matrices on the support. For commuting generators $U_i$ of orders $n_i$, the same projection can be performed one generator at a time:
\begin{equation}
\Pi_{\bm q}(O)=\prod_i\left[\frac1{n_i}\sum_{s=0}^{n_i-1}e^{-2\pi\ii q_i s/n_i}U_i^sOU_i^{-s}\right],
\label{eq:SM_generator_projector}
\end{equation}
where $\bm q=(q_1,q_2,\ldots)$ labels the symmetry charges. Here $\rho$ is the stationary, symmetry-preserving density matrix that defines the operator inner product in Eq.~\eqref{eq:SM_inner}; for the thermal calculations it is the Gibbs state $\rho_\beta$. Its symmetry invariance, $U_g\rho U_g^\dagger=\rho$, makes operators belonging to distinct character sectors orthogonal in $C$. Since $[H,U_g]=0$, commutation with $H$ does not change an operator's symmetry charge, so the sectors are also orthogonal in $D$. Consequently $C$ and $D$, and hence the generalized eigenproblem, are exactly block diagonal in the charge sectors. Projecting only after diagonalization can mix eigenvectors from tied charge sectors and thereby lose this charge information.

\subsection{Local reduced-state evaluation of $C$ and $D$}
\label{subsec:SM_local_reduced_state}

Equation~\eqref{eq:SM_CD} is written as a trace over the full chain, but a finite-range Hamiltonian makes its actual input local. Let $A$ denote the $\ell$-site classification window and define its \emph{interaction collar}
\begin{equation}
	\mathcal B_A=
	\operatorname{hull}\!\left(
	A\cup\bigcup_{\substack{X:\,h_X\neq0\\X\cap A\neq\varnothing}}X
	\right).
	\label{eq:SM_interaction_collar}
\end{equation}
Here $\operatorname{hull}$ means the smallest lattice interval containing the indicated set. Thus $\mathcal B_A$ contains $A$ and every site belonging to a Hamiltonian term that touches $A$. Equation~\eqref{eq:SM_finite_range_H} implies $\mathcal B_A\subseteq A^{[R]}$. For a basis operator supported on $A$,
\begin{equation}
	Q_a\equiv[H,O_a]
	=\sum_{X\cap A\neq\varnothing}[h_X,O_a],
	\qquad
	\operatorname{supp}Q_a\subseteq\mathcal B_A,
	\label{eq:SM_local_commutator}
\end{equation}
because every term disjoint from $A$ commutes with $O_a$. For example, if $A=\{0,\ldots,\ell-1\}$, a nearest-neighbor Hamiltonian has $\mathcal B_A\subseteq\{0,\ldots,\ell\}$, whereas the three-site cluster and clock terms can extend the collar through site $\ell+1$.

\begin{corollary}[Local thermal input for classification]
	\label{cor:SM_local_classifier}
	Let $H$ satisfy Eq.~\eqref{eq:SM_finite_range_H}, take $\rho=\rho_{\beta,L}$ in Eq.~\eqref{eq:SM_inner}, let every $O_a$ be supported on $A$, and write
	\begin{equation}
		\rho_Y=\Tr_{\Lambda_L\setminus Y}(\rho_{\beta,L}),
		\qquad
		\delta_AO_a=O_a-\Tr_A(\rho_AO_a)\mathds1_A.
		\label{eq:SM_local_marginals}
	\end{equation}
	Then the full-chain matrices in Eq.~\eqref{eq:SM_CD} are exactly
	\begin{align}
		C_{ab}&=\frac12\Tr_A\!\left[
		\rho_A\left(\delta_AO_a^\dagger\delta_AO_b+
		\delta_AO_b\delta_AO_a^\dagger\right)\right],
		\label{eq:SM_C_from_rhoA}\\
		D_{ab}&=\frac12\Tr_{\mathcal B_A}\!\left[
		\rho_{\mathcal B_A}\left(Q_a^\dagger Q_b+Q_bQ_a^\dagger\right)\right].
		\label{eq:SM_D_from_rhoB}
	\end{align}
	Consequently, $C$ requires only the Gibbs marginal on $A$, and $D$ requires only the marginal on $\mathcal B_A$.
\end{corollary}

\noindent\textit{Proof.} The identity for $C$ follows directly by tracing out the complement of $A$. For $D$, stationarity removes connected subtraction:
\begin{equation}
	\Tr(\rho_{\beta,L}Q_a)
	=\Tr\!\left([\rho_{\beta,L},H]O_a\right)=0.
	\label{eq:SM_commutator_mean_zero}
\end{equation}
Equation~\eqref{eq:SM_local_commutator} then allows every site outside $\mathcal B_A$ to be traced out, which gives Eq.~\eqref{eq:SM_D_from_rhoB}. \hfill$\square$

At $\beta=0$, $\rho_Y=\mathds1_Y/d^{|Y|}$ for every $Y$. Equations~\eqref{eq:SM_C_from_rhoA} and \eqref{eq:SM_D_from_rhoB} are therefore exact normalized traces on $A$ and $\mathcal B_A$, with no Hamiltonian diagonalization and no dependence on $L$ once the collar fits.

At finite $\beta$, the exact marginals can be approximated without constructing the Gibbs state of the full chain. Choose an interval $W\supseteq\mathcal B_A$, keep only the Hamiltonian terms contained in it, and form
\begin{equation}
	H_W=\sum_{X\subseteq W}h_X,
	\qquad
	\sigma_W=\frac{e^{-\beta H_W}}{\Tr_W(e^{-\beta H_W})},
	\qquad
	\sigma_Y^{(W)}=\Tr_{W\setminus Y}(\sigma_W).
	\label{eq:SM_guarded_Gibbs}
\end{equation}
The sites between $\mathcal B_A$ and an artificial boundary of $W$ form the \emph{guard}; set
\begin{equation}
	s=\operatorname{dist}(\mathcal B_A,\Lambda_L\setminus W).
	\label{eq:SM_guard_width}
\end{equation}
For a left-edge calculation, $W$ begins at the physical left end and only a right guard is needed. For a bulk or periodic calculation, the target region must have a guard on both sides. Since every term contributing to $Q_a$ lies inside $\mathcal B_A\subseteq W$, the same local commutator is obtained from $H_W$: $[H_W,O_a]=[H,O_a]$.

Let $C^{(W)}$ and $D^{(W)}$ be Eqs.~\eqref{eq:SM_C_from_rhoA} and \eqref{eq:SM_D_from_rhoB} with $\rho_A,\rho_{\mathcal B_A}$ replaced by $\sigma_A^{(W)},\sigma_{\mathcal B_A}^{(W)}$. Define
\begin{equation}
	\epsilon_W=\left\|\rho_{\mathcal B_A}-
	\sigma_{\mathcal B_A}^{(W)}\right\|_1.
	\label{eq:SM_guard_error}
\end{equation}
For bounded finite-range interactions in one dimension, finite-temperature local indistinguishability gives, at fixed $\beta$, $A$, and uniform interaction bounds~\cite{Capel2025Gibbs},
\begin{equation}
	\epsilon_W\leq K_{\beta,A}(1+\sqrt{s})e^{-c_\beta\sqrt{s}},
	\qquad 0<\beta<\infty,
	\label{eq:SM_guard_stretched}
\end{equation}
where $K_{\beta,A}<\infty$ and $c_\beta>0$ are independent of $L$ and $W$. Translation invariance is not required for this bound. For a translation-invariant or fixed-period finite-range chain, the stronger one-dimensional Gibbs estimates sharpen Eq.~\eqref{eq:SM_guard_stretched} to $\epsilon_W\leq K_{\beta,A}e^{-\gamma_\beta s}$ for some $\gamma_\beta>0$~\cite{Araki1969Gibbs,Capel2025Gibbs}. At $\beta=0$, $\epsilon_W=0$ as soon as $W$ contains $\mathcal B_A$.

These state bounds control the classifier directly. Contractivity of trace distance under partial trace and $|\Tr[(\rho-\sigma)M]|\leq\|\rho-\sigma\|_1\|M\|$ give
\begin{align}
	|C_{ab}^{(W)}-C_{ab}|&\leq
	3\epsilon_W\|O_a\|\|O_b\|,
	\label{eq:SM_C_guard_bound}\\
	|D_{ab}^{(W)}-D_{ab}|&\leq
	\epsilon_W\|Q_a\|\|Q_b\|.
	\label{eq:SM_D_guard_bound}
\end{align}
The factor three in Eq.~\eqref{eq:SM_C_guard_bound} allows for the change of the two one-point functions in the connected covariance; it can be improved in symmetry sectors whose one-point functions vanish. At fixed $\ell$, the basis is finite, so $C^{(W)}\to C$ and $D^{(W)}\to D$ in every matrix norm. If the retained part of $C$ stays bounded away from zero and the desired eigenspace of $C^{-1/2}DC^{-1/2}$ is separated from the rest by a nonzero gap, standard finite-dimensional perturbation theory then gives convergence of its generalized eigenvalues and spectral projector at the same asymptotic rate. If the lowest eigenvalue is degenerate, the complete tied space must be retained as in Sec.~\ref{sec:SM_degenerate}; individual eigenvectors inside it need not converge.

The practical workflow is therefore:
\begin{enumerate}
	\item Choose $A$ and the symmetry-resolved basis $\{O_a\}$.
	\item Build $\mathcal B_A$ and $Q_a$ from only the Hamiltonian terms that touch $A$.
	\item At $\beta=0$, evaluate the normalized traces on $A$ and $\mathcal B_A$. At finite $\beta$, choose $W\supseteq\mathcal B_A$, form $\sigma_W$, and trace it down to the two required regions.
	\item Enlarge only the guard and repeat until $C$, $D$, the relevant tied eigenspaces, and the final charge or algebra test are stable within the requested tolerance.
\end{enumerate}
For the nearest-neighbor AKLT and large-$D$ chains, the interaction collar adds at most one site to a left boundary window. For the cluster and alternating clock Hamiltonians, it adds at most two sites; because their fixed-point calculations use $\beta=0$, no guard is needed after that collar fits. The same construction applies to detection at fixed Krylov depth $K$: every raw iterate through $Q_{2K}$ lies in the commutator cone $\Gamma_K=Y_0^{[2KR]}$, so a guarded Gibbs calculation on an interval containing $\Gamma_K$ determines the finite Gram matrix in Eq.~\eqref{eq:SM_raw_Gram}. The required core grows with $K$, as it must; the claim is independence from the total chain length at fixed $K$, not a fixed-size calculation as $K\to\infty$.

For fixed $\ell$ and target matrix accuracy $\varepsilon$, the size of $W$ is independent of $L$: Eq.~\eqref{eq:SM_guard_stretched} permits $s=O(\log^2(1/\varepsilon))$, while the fixed-period exponential bound permits $s=O(\log(1/\varepsilon))$. If $H_W$ is treated by dense methods, the Hilbert-space dimension is $d^{|W|}$ rather than $d^L$. This removes full-chain exact diagonalization, but it does not remove the exponential dependence on the chosen local window, guard, or Krylov depth.

The finite-range hypothesis is essential to this system-size reduction. With genuinely long-range terms, $\mathcal B_A$ can equal the full chain even when $A$ is small. The identities in Corollary~\ref{cor:SM_local_classifier} remain true, but then they provide no computational saving. If the interactions decay with distance, let $H^{(r)}$ retain, among the terms that touch $A$, only those supported in $A^{[r]}$. For every $O$ supported on $A$, truncating the remaining terms gives the explicit tail error
\begin{equation}
	\left\|[H-H^{(r)},O]\right\|
	\leq2\|O\|
	\sum_{\substack{X\cap A\neq\varnothing\\X\nsubseteq A^{[r]}}}\|h_X\|.
	\label{eq:SM_long_range_tail}
\end{equation}
This controls only the commutator; approximating the Gibbs marginal also requires an appropriate long-range local-indistinguishability bound. No system-size-independent local algorithm is claimed here without quantitative decay assumptions that control both errors.

\subsection{Anchored endpoint spaces and class readout}
\label{subsec:SM_anchor_readout}

In a one-dimensional bosonic SPT phase, the action of a global symmetry on the low-energy space of a long open chain separates into contributions localized near the two ends. If $g$ is an element of the protecting symmetry group and $U_g$ is its global many-body symmetry operator, then
\begin{equation}
	U_g\simeq V_g^{\rm L}V_g^{\rm R},
	\qquad
	V_g^{\rm L}V_h^{\rm L}=e^{\ii\phi(g,h)}V_{gh}^{\rm L},
	\label{eq:SM_fractionalization}
\end{equation}
up to quasi-local dressing and exponentially small coupling between the two ends~\cite{Chen2011prb,Pollmann2010,Pollmann2012}. Here $V_g^{\rm L}$ and $V_g^{\rm R}$ are the symmetry actions localized near the left and right boundaries, respectively, while $V_{gh}^{\rm L}$ is the left endpoint associated with the product group element $gh$. The symbol $\simeq$ denotes equality within the low-energy open-chain space up to the stated finite-size and quasi-local corrections. The phase $e^{\ii\phi(g,h)}$ means that the endpoint operators can realize the symmetry only projectively: multiplying two endpoint actions can differ from the endpoint associated with $gh$ by a phase. For the clock models this phase appears as $A_1A_2=\omega_N^pA_2A_1$, while for the AKLT edge the two $D_2$ endpoint actions anticommute.

The classification target is therefore a specific symmetry endpoint $V_g^{\rm L}$, rather than an arbitrary conserved operator that happens to be localized near the boundary. Anchoring enforces this distinction: the microscopic left factor of a chosen global symmetry generator is held fixed, while additional sites are added only to dress that factor as the boundary window grows. Thus increasing $\ell$ improves the quasi-local representation of the same endpoint instead of allowing the variational search to switch to an unrelated conserved boundary operator.

For $G=\mathbb Z_N\times\mathbb Z_N$, $\mathbb Z_N$ is the cyclic group of integers modulo $N$. In the clock chain, let $U_1=\prod_{j\ {\rm even}}X_j$ and $U_2=\prod_{j\ {\rm odd}}X_j$. The first endpoint is anchored by the leftmost local factor $X_0$ of $U_1$, while every additional basis operator begins on site one:
\begin{equation}
\mathcal S_{1,\ell}=\operatorname{span}\left\{X_0\prod_{j=1}^{\ell-1}X_j^{a_j}Z_j^{b_j}:a_j,b_j=0,\ldots,N-1\right\}.
\label{eq:SM_anchor_space}
\end{equation}
For example, at $\ell=3$ a basis element is $X_0X_1^{a_1}Z_1^{b_1}X_2^{a_2}Z_2^{b_2}$. The second endpoint is anchored by the leftmost local factor $X_1$ of $U_2$, so at the same window size a basis element has the form $X_1X_2^{a_2}Z_2^{b_2}$. This offset keeps the fixed symmetry factor separate from the dressing variables. The Hilbert-space dimension of an $\ell$-site clock window is $d_\partial=N^\ell$.
The charge-$q$ projector under $U_2$ is
\begin{equation}
\Pi_q(O)=\frac1N\sum_{s=0}^{N-1}\omega_N^{-qs}U_2^sOU_2^{-s}.
\label{eq:SM_q_projector}
\end{equation}
Solve Eq.~\eqref{eq:SM_generalized_problem} separately for $q=0,\ldots,N-1$. The recovered first endpoint is denoted $A_1^\ell$. Recover $A_2^\ell$ independently by anchoring the leftmost microscopic factor $X_1$ of $U_2$. This second recovery is a redundancy check rather than the primary class readout.

Suppose the exact left endpoints satisfy
\begin{equation}
A_1A_2=\omega_N^pA_2A_1.
\label{eq:SM_projective_algebra}
\end{equation}
Conjugating $A_1$ by $A_2$ gives $A_2A_1A_2^\dagger=\omega_N^{-p}A_1$. The remote right endpoint commutes with $A_1$, so the global second symmetry has the same action:
\begin{equation}
U_2A_1U_2^\dagger=\omega_N^qA_1,
\qquad
q=-p\pmod N.
\label{eq:SM_p_readout}
\end{equation}
Therefore the charge of $A_1$, not $A_2$, gives the primary class label. The independently recovered pair must also satisfy Eq.~\eqref{eq:SM_projective_algebra} within its controlled residual.

These checks are direct matrix tests. On a common boundary Hilbert space of dimension $d_\partial$, define $\norm{M}_{\rm F,n}=\norm{M}_{\rm F}/\sqrt{d_\partial}$. If $u_2^\partial$ is the restriction of $U_2$ to that window, use
\begin{equation}
\epsilon_q(A_1)=\frac{\norm{u_2^\partial A_1u_2^{\partial\dagger}-\omega_N^qA_1}_{\rm F,n}}{\norm{A_1}_{\rm F,n}},
\qquad
\epsilon_p(A_1,A_2)=\norm{A_1A_2-\omega_N^pA_2A_1}_{\rm F,n}.
\label{eq:SM_charge_algebra_residuals}
\end{equation}
Both residuals vanish for an exact endpoint pair. For $D_2$, whose variational endpoints are controlled only where the state $\rho$ has weight, the nontrivial spin-$1/2$ edge relation is tested in the state norm of Eq.~\eqref{eq:SM_inner}, $\norm{A_xA_z+A_zA_x}_S/(\norm{A_x}_S\norm{A_z}_S)$, while a commuting pair uses the minus sign; the normalized Frobenius norm weights all window states equally and does not become small for these endpoints.

\begin{proposition}[Convergence of a complete anchored search]
\label{prop:SM_classification_convergence}
Assume that a normalized endpoint $A_1$ of charge $q$ has truncations $A_1^{(\ell)}\in\mathcal S_{1,\ell}^{(q)}$ satisfying
\begin{equation}
\norm{A_1-A_1^{(\ell)}}_S+\norm{[H,A_1-A_1^{(\ell)}]}_S\leq Ce^{-\ell/\xi}.
\label{eq:SM_graph_locality}
\end{equation}
If $[H,A_1]=0$, then the lowest stiffness in that anchored charge sector obeys $\kappa_{1,q}(\ell)\leq C'e^{-2\ell/\xi}$. If the endpoint obeys $\norm{[H,A_1]}_S\leq\delta$, then $\limsup_{\ell\to\infty}\kappa_{1,q}(\ell)\leq\delta^2$.
\end{proposition}

\noindent\textit{Proof.} Use the truncation $A_1^{(\ell)}\in\mathcal S_{1,\ell}^{(q)}$ itself as a trial operator in the Rayleigh quotient of Eq.~\eqref{eq:SM_Rayleigh}. Since $A_1$ is normalized, Eq.~\eqref{eq:SM_graph_locality} gives
\begin{equation}
	\norm{A_1^{(\ell)}}_S
	\geq
	\norm{A_1}_S-\norm{A_1-A_1^{(\ell)}}_S
	\geq
	1-Ce^{-\ell/\xi},
\end{equation}
so the denominator of the Rayleigh quotient approaches one and is nonzero for sufficiently large $\ell$.

First suppose that $A_1$ is an exact endpoint, $[H,A_1]=0$. Then
\begin{equation}
	\norm{[H,A_1^{(\ell)}]}_S
	=
	\norm{[H,A_1^{(\ell)}-A_1]}_S
	\leq
	Ce^{-\ell/\xi}.
\end{equation}
Therefore the Rayleigh quotient of this particular trial operator obeys
\begin{equation}
	\frac{\norm{[H,A_1^{(\ell)}]}_S^2}{\norm{A_1^{(\ell)}}_S^2}
	\leq
	\frac{C^2e^{-2\ell/\xi}}{\left(1-Ce^{-\ell/\xi}\right)^2}
	=
	O(e^{-2\ell/\xi}).
\end{equation}
Because $\kappa_{1,q}(\ell)$ is the minimum of the Rayleigh quotient over the entire anchored charge sector, it cannot exceed the value obtained from this trial operator. This proves $\kappa_{1,q}(\ell)\leq C'e^{-2\ell/\xi}$ for sufficiently large $\ell$.

If instead $\norm{[H,A_1]}_S\leq\delta$, the triangle inequality gives
\begin{equation}
	\norm{[H,A_1^{(\ell)}]}_S
	\leq
	\norm{[H,A_1]}_S+\norm{[H,A_1^{(\ell)}-A_1]}_S
	\leq
	\delta+Ce^{-\ell/\xi}.
\end{equation}
Hence
\begin{equation}
	\kappa_{1,q}(\ell)
	\leq
	\frac{\left(\delta+Ce^{-\ell/\xi}\right)^2}
	{\left(1-Ce^{-\ell/\xi}\right)^2},
\end{equation}
and taking $\ell\to\infty$ gives $\limsup_{\ell\to\infty}\kappa_{1,q}(\ell)\leq\delta^2$. \hfill$\square$

This proposition explains why a growing complete anchored space finds a quasi-local endpoint. A stable phase label additionally requires isolation from wrong-charge branches, open-versus-periodic contrast, and convergence of the recovered subspace. For a charge sector $\chi$, define
\begin{equation}
\eta_\chi(\ell)=\frac{\kappa_{1,\chi}^{\rm PBC}(\ell)}{\kappa_{1,\chi}^{\rm OBC}(\ell)}.
\label{eq:SM_eta}
\end{equation}
A clean boundary branch has small OBC stiffness while the corresponding PBC stiffness remains on a generic bulk scale, so $\eta_\chi(\ell)\gg1$. Because the search spaces are nested, Eq.~\eqref{eq:SM_kappa_monotone} implies that each lowest stiffness can only stay fixed or decrease as $\ell$ grows. The useful finite-size warning therefore comes from the PBC branch: if enlarging the window from $\ell$ to $\ell+1$ lets the periodic search access an additional slow structure, $\kappa_{1,\chi}^{\rm PBC}$ can drop sharply and $\eta_\chi$ can collapse. When the charge and recovered endpoint are already stable, we take $\ell$ as the last clean boundary window before such a sharp drop. This is a practical finite-size selection rule, while Eq.~\eqref{eq:SM_kappa_monotone} is the exact variational statement. If $\kappa_{1,\chi}^{\rm OBC}=0$, the open chain has an exact zero-stiffness operator in that space and the ratio is recorded separately.

When the retained branch has dimension greater than one, individual eigenvectors can rotate arbitrarily as $\ell$ changes. Let $\{A_i^{(\ell)}\}$ and $\{A_j^{(\ell+1)}\}$ be state-metric orthonormal bases of the two retained subspaces. To compare them on the same Hilbert space, embed an operator from the $\ell$-site window into the $(\ell+1)$-site window by tensoring it with the identity on the newly added site. We denote this inclusion by $\iota$; for example, in the clock chain $\iota A^{(\ell)}=A^{(\ell)}\otimes\mathds 1_N$. Form
\begin{equation}
M_{ji}^{(\ell)}=(A_j^{(\ell+1)}|\iota A_i^{(\ell)})_S.
\label{eq:SM_subspace_overlap}
\end{equation}
The singular values of $M^{(\ell)}$ lie between zero and one. A value one means that a direction in the smaller retained space is reproduced exactly in the larger one, while a value near zero means that the corresponding directions have little overlap. Singular values approaching one therefore show window stability without depending on which linear combinations the eigensolver returns inside a degenerate subspace.

For the symmetry-preserving boundary perturbation $H(h_\partial)=H_0+h_\partial V_\partial$ used in the Letter and an exact endpoint $A_0$ already contained in the search space,
\begin{equation}
\kappa_1(h_\partial)\leq h_\partial^2\frac{\norm{[V_\partial,A_0]}_{S,h_\partial}^2}{\norm{A_0}_{S,h_\partial}^2}.
\label{eq:SM_perturbative_bound}
\end{equation}
This follows by inserting $A_0$ into Eq.~\eqref{eq:SM_Rayleigh}. The optimized endpoint may dress within the boundary window and have an even smaller stiffness. In a fixed finite search space, $C$ and $D$ vary continuously with $h_\partial$, so a branch separated from wrong-charge sectors remains separated for sufficiently small $|h_\partial|$.

\subsection{Exact alternating clock fixed-point representative for every $N\geq2$}
\label{subsec:SM_clock_fixed_point}

Here ``fixed point'' means an exactly solvable representative of the SPT phase built from mutually commuting local terms, with exact boundary operators of finite support. It is a phase representative, not a critical point. For these operators the zero-frequency structure is already exact at $\beta=0$, while finite $\beta$ remains an allowed choice of the metric.

Define $\epsilon_j=(-1)^{j+1}$ and
\begin{equation}
K_j^{(p)}=Z_{j-1}^{\epsilon_jp}X_jZ_{j+1}^{-\epsilon_jp},
\qquad
p\in\mathbb Z_N.
\label{eq:SM_clock_stabilizer}
\end{equation}
The open and periodic Hamiltonians are
\begin{align}
H_p^{\rm OBC}&=-\frac12\sum_{j=1}^{L-2}\left(K_j^{(p)}+K_j^{(p)\dagger}\right),
\qquad L\geq3,\nonumber\\
H_p^{\rm PBC}&=-\frac12\sum_{j=0}^{L-1}\left(K_j^{(p)}+K_j^{(p)\dagger}\right),
\qquad L\ {\rm even},
\label{eq:SM_clock_H}
\end{align}
where periodic labels are taken modulo $L$. Open chains allow either parity of $L$. The construction and all conclusions below hold for odd and even $N$.

Nonadjacent stabilizers commute. Adjacent stabilizers acquire one Weyl phase from each shared bond, and $\epsilon_{j+1}=-\epsilon_j$ makes the two phases cancel. Thus
\begin{equation}
[K_j^{(p)},K_k^{(p)}]=0,
\qquad
[K_j^{(p)},U_1]=[K_j^{(p)},U_2]=0.
\label{eq:SM_clock_commuting}
\end{equation}
To expose the spectrum, introduce the two-qudit controlled-phase gate on neighboring sites $j$ and $j+1$. In the computational basis $|a,b\rangle\equiv|a\rangle_j\otimes|b\rangle_{j+1}$, with $a,b=0,\ldots,N-1$ and $\omega_N=e^{2\pi\ii/N}$,
\begin{equation}
	CZ_{j,j+1}
	=
	\sum_{a,b=0}^{N-1}\omega_N^{ab}|a,b\rangle\langle a,b|,
	\qquad
	CZ_{j,j+1}|a,b\rangle=\omega_N^{ab}|a,b\rangle.
	\label{eq:SM_CZ}
\end{equation}
Thus $CZ$ is diagonal in the computational basis and changes only the phase of each basis state. For $N=2$, where $\omega_2=-1$, this reduces to the familiar qubit gate
\begin{equation}
	CZ=\operatorname{diag}(1,1,1,-1)
\end{equation}
in the ordered basis $|00\rangle,|01\rangle,|10\rangle,|11\rangle$.

The alternating clock fixed point is related to a decoupled onsite Hamiltonian by
\begin{equation}
	\mathcal U_p=\prod_jCZ_{j,j+1}^{(-1)^jp}.
	\label{eq:SM_clock_circuit}
\end{equation}
The exponent means that neighboring bonds alternately carry powers $CZ^p$ and $CZ^{-p}$; negative powers are inverse gates. All $CZ$ gates are diagonal and commute, and the nearest-neighbor circuit can be implemented in two layers, one on even bonds and one on odd bonds. Its depth is therefore independent of $L$.
With the appropriate open or even-periodic bond set,
\begin{equation}
\mathcal U_pX_j\mathcal U_p^\dagger=K_j^{(p)}.
\label{eq:SM_clock_circuit_action}
\end{equation}
Therefore $H_p^{\rm OBC}$ is unitarily equivalent to $-\frac12\sum_{j=1}^{L-2}(X_j+X_j^\dagger)$, while $H_p^{\rm PBC}$ is equivalent to the same onsite sum over every site. Because unitary conjugation does not change eigenvalues, the spectrum can be read from these decoupled one-site terms. Let $|\widetilde r\rangle$, $r=0,\ldots,N-1$, be an eigenstate of $X$,
\begin{equation}
	X|\widetilde r\rangle=\omega_N^r|\widetilde r\rangle.
\end{equation}
For the one-site Hamiltonian $h=-\frac12(X+X^\dagger)$,
\begin{equation}
	h|\widetilde r\rangle
	=
	-\frac12\left(\omega_N^r+\omega_N^{-r}\right)|\widetilde r\rangle
	=
	-\cos\left(\frac{2\pi r}{N}\right)|\widetilde r\rangle.
\end{equation}
The minimum occurs at $r=0$, with energy $E_0=-1$. The lowest excited one-site level is $r=1$ or equivalently $r=N-1$, with energy $E_1=-\cos(2\pi/N)$. Since the transformed many-body Hamiltonian is a sum of independent onsite terms, the smallest excitation above the ground sector is obtained by exciting one such site. The exact many-body gap is therefore
\begin{equation}
	\Delta_N=E_1-E_0
	=1-\cos\frac{2\pi}{N}>0.
	\label{eq:SM_clock_gap}
\end{equation}
The $L-2$ open stabilizers are independent because the power of the unique central $X_j$ fixes every exponent in a product relation. Hence the open ground space has dimension $N^2$. The even-periodic problem has $L$ independent stabilizers and a unique ground state.

A convenient exact first left endpoint is
\begin{equation}
A_1=X_0Z_1^p.
\label{eq:SM_clock_A1_exact}
\end{equation}
Only $K_1^{(p)}$ overlaps nontrivially with $A_1$. Its Weyl phase at site zero cancels the opposite phase at site one, proving $[H_p^{\rm OBC},A_1]=0$. A second exact representative is $Z_0^{-p}$. The classifier, however, anchors the second endpoint with the leftmost microscopic factor $X_1$ of $U_2$, so the directly comparable representative is
\begin{equation}
A_2=X_1Z_2^{-p}=Z_0^{-p}K_1^{(p)}.
\label{eq:SM_clock_endpoints}
\end{equation}
Because both $Z_0^{-p}$ and $K_1^{(p)}$ commute with $H_p^{\rm OBC}$, this anchored $A_2$ also commutes exactly with the Hamiltonian. The two anchored endpoints satisfy
\begin{equation}
A_1A_2=\omega_N^pA_2A_1,
\qquad
U_2A_1U_2^\dagger=\omega_N^{-p}A_1.
\label{eq:SM_clock_class}
\end{equation}
Thus $A_1$ supplies the primary charge readout $q=-p\pmod N$, while $A_2$ supplies an independent algebra check. Inside the stabilizer ground space, $A_2$ and $Z_0^{-p}$ act identically because $K_1^{(p)}=1$.

The algebra in Eq.~\eqref{eq:SM_clock_class} also determines the dimension of the irreducible projective degree of freedom carried by one edge. Define
\begin{equation}
	d_p=\frac{N}{\gcd(N,p)}.
\end{equation}
Equivalently, $d_p$ is the smallest positive integer for which
\begin{equation}
	\left(\omega_N^p\right)^{d_p}=1.
\end{equation}
This is what is meant by saying that the phase $\omega_N^p$ has order $d_p$.

To see why $d_p$ is an edge dimension, let $|v\rangle$ be an eigenvector of $A_2$, with $A_2|v\rangle=\lambda|v\rangle$. From $A_1A_2=\omega_N^pA_2A_1$,
\begin{equation}
	A_2A_1^r|v\rangle
	=
	\lambda\left(\omega_N^p\right)^{-r}A_1^r|v\rangle.
\end{equation}
For $r=0,\ldots,d_p-1$, these eigenvalues are distinct, so the vectors $|v\rangle,A_1|v\rangle,\ldots,A_1^{d_p-1}|v\rangle$ are linearly independent. After $d_p$ steps the phase returns to one and the orbit closes. Thus any representation of this projective algebra contains a $d_p$-dimensional factor, and the ordinary $d_p$-dimensional clock and shift matrices realize such an irreducible factor explicitly.

The number $d_p$ is therefore the protected projective edge dimension, not necessarily the full microscopic edge degeneracy of the fixed-point Hamiltonian. For example, if $p$ is coprime to $N$, then $d_p=N$; for $N=4$ and $p=2$, $d_p=2$; and for $p=0$, $d_p=1$, so there is no nontrivial protected projective factor. When $p$ and $N$ share a divisor, the larger fixed-point edge space can contain additional reducible multiplicity. Symmetry-preserving boundary terms may split that multiplicity, but they cannot remove the irreducible $d_p$-dimensional projective factor.

For $p=0$, the Hamiltonian omits the endpoint variables and produces an accidental free-edge response. The symmetry-preserving boundary perturbation
\begin{equation}
V_\partial=\frac12\left(X_0+X_0^\dagger+X_{L-1}+X_{L-1}^\dagger\right)
\label{eq:SM_p_zero_pin}
\end{equation}
fixes these variables while preserving $U_1$ and $U_2$. For $p\neq0$, the projective factor in Eq.~\eqref{eq:SM_clock_class} remains protected. Section~\ref{subsec:SM_p_zero_degeneracy} uses $N=3,p=0$ as an exactly soluble illustration of accidental degeneracy. The general recovery rule in Sec.~\ref{sec:SM_degenerate} applies equally when a within-sector degeneracy or a tie between charge sectors occurs at nonzero $p$.

\subsection{AKLT as an exact projective-representation example}
\label{subsec:SM_AKLT_projective}

The spin-one AKLT Hamiltonian is
\begin{equation}
H_{\rm AKLT}=\sum_{j=0}^{L-2}\left[\boldsymbol S_j\cdot\boldsymbol S_{j+1}+\frac13\left(\boldsymbol S_j\cdot\boldsymbol S_{j+1}\right)^2\right].
\label{eq:SM_AKLT_H}
\end{equation}
An irrelevant additive constant has been omitted from Eq.~\eqref{eq:SM_AKLT_H}. The comparison chain adds a positive single-ion anisotropy,
\begin{equation}
H_{\rm LD}=H_{\rm AKLT}+D\sum_{j=0}^{L-1}(S_j^z)^2,
\qquad D>0.
\label{eq:SM_large_D_H}
\end{equation}
As $D\to\infty$, the ground state approaches the product state $|0\rangle^{\otimes L}$. The sufficiently large-$D$ phase is topologically trivial while preserving the same $D_2$ symmetry.
The AKLT ground space has a particularly transparent matrix-product-state representation. For an open chain, a ground state can be written as
\begin{equation}
	|\Psi(w_{\rm L},w_{\rm R})\rangle
	=
	\sum_{m_0,\ldots,m_{L-1}}
	w_{\rm L}^\dagger
	M^{m_0}M^{m_1}\cdots M^{m_{L-1}}
	w_{\rm R}
	\,|m_0,m_1,\ldots,m_{L-1}\rangle,
\end{equation}
where each physical index $m_j\in\{+1,0,-1\}$ labels a spin-one basis state, while the $M^m$ are $2\times2$ matrices acting on a two-dimensional auxiliary, or ``virtual,'' space. The boundary vectors $w_{\rm L}$ and $w_{\rm R}$ select states in the left and right virtual edge spaces. This two-dimensional virtual space is the spin-$1/2$ edge degree of freedom of the AKLT chain. One convenient gauge is
\begin{equation}
	M^{+1}=\sqrt{\frac23}\,\sigma^+,
	\qquad
	M^0=-\frac1{\sqrt3}\sigma^z,
	\qquad
	M^{-1}=-\sqrt{\frac23}\,\sigma^-.
	\label{eq:SM_AKLT_MPS}
\end{equation}
Direct substitution of the physical matrices in Eq.~\eqref{eq:SM_D2_matrices} gives the intertwining relations
\begin{equation}
\sum_n(u_x)_{mn}M^n=\sigma^xM^m\sigma^x,
\qquad
\sum_n(u_z)_{mn}M^n=\sigma^zM^m\sigma^z.
\label{eq:SM_AKLT_intertwining}
\end{equation}
The virtual endpoint actions obey
\begin{equation}
\sigma^x\sigma^z=-\sigma^z\sigma^x.
\label{eq:SM_AKLT_anticommutation}
\end{equation}
Thus the globally commuting $D_2$ half-turns realize the nontrivial projective class on the AKLT edge~\cite{AKLT1987,Pollmann2010,Pollmann2012}. In the large-$D$ product limit, the state is built from $|0\rangle$ and the virtual space is one dimensional, so the endpoint symmetry phases commute. This gives the analytic distinction that the recovered three nontrivial $D_2$ sectors are designed to test.

\subsection{Spin-one $D_2$ boundary search spaces}
\label{subsec:SM_spin1_spaces}

For the spin-one classifier, use the complete $D_2$-resolved operator space on the first $\ell$ sites. This differs from the anchored clock construction because no microscopic factor such as $X_0$ is fixed. Let $B_\alpha$ run over the identity together with the eight traceless one-site operators in Eq.~\eqref{eq:SM_spin_one_bank}, and let $\bm q_\alpha=(q_{x,\alpha},q_{z,\alpha})\in\mathbb Z_2^2$ denote its $D_2$ charge. A tensor-product operator has the charge obtained by adding its local charge bits modulo two:
\begin{equation}
	O_{\bm\alpha}=B_{\alpha_0}\otimes\cdots\otimes B_{\alpha_{\ell-1}},
	\qquad
	\bm q(O_{\bm\alpha})=\sum_{j=0}^{\ell-1}\bm q_{\alpha_j}\pmod 2.
	\label{eq:SM_spin1_charge_sum}
\end{equation}
The complete boundary search space in sector $\bm q$ is therefore
\begin{equation}
	\mathcal S_\ell^{(\bm q)}
	=
	\operatorname{span}\left\{
	B_{\alpha_0}\otimes\cdots\otimes B_{\alpha_{\ell-1}}:
	\sum_{j=0}^{\ell-1}\bm q_{\alpha_j}=\bm q\pmod 2
	\right\}.
	\label{eq:SM_spin1_sector_space}
\end{equation}
Equation~\eqref{eq:SM_spin1_charge_sum} is the spin-one analogue of the Weyl-exponent constraints in the clock construction: a basis string is retained precisely when its two accumulated charge bits equal the requested sector. Tensoring an existing operator with $\mathds 1$ preserves its charge, so $\mathcal S_\ell^{(\bm q)}\subseteq\mathcal S_{\ell+1}^{(\bm q)}$.

Including the identity, the one-site neutral sector has dimension three and each nontrivial sector has dimension two. Hence
\begin{equation}
	\dim\mathcal S_\ell^{(0,0)}=\frac{9^\ell+3}{4},
	\qquad
	\dim\mathcal S_\ell^{(\bm q\neq(0,0))}=\frac{9^\ell-1}{4}.
	\label{eq:SM_spin1_sector_dimensions}
\end{equation}
For the first five windows,
\begin{equation}
	\begin{array}{c|ccccc}
		\ell&1&2&3&4&5\\
		\hline
		\dim\mathcal S_\ell^{(0,0)}&3&21&183&1641&14763\\
		\dim\mathcal S_\ell^{(\bm q\neq(0,0))}&2&20&182&1640&14762
	\end{array}
	\label{eq:SM_spin1_sector_dimension_table}
\end{equation}
before state-null directions are removed. For example, at $\ell=1$ the $(0,1)$ sector is spanned by $S_0^x$ and $S_0^yS_0^z+S_0^zS_0^y$. At $\ell=2$, the same target sector contains all tensor products whose local charges are $(0,0)+(0,1)$, $(0,1)+(0,0)$, $(1,0)+(1,1)$, or $(1,1)+(1,0)$. The same modulo-two rule generates the basis automatically at arbitrary $\ell$. The generalized eigenproblem in Eq.~\eqref{eq:SM_generalized_problem} is solved independently in the three nontrivial sectors $(0,1)$, $(1,0)$, and $(1,1)$. Their slow operators are denoted $A_x^\ell$, $A_z^\ell$, and $A_y^\ell$, respectively. The practical class-assignment rule is given in Sec.~\ref{subsec:SM_spin1_final_rule}.

\section{Degenerate minima and certified recovery of the left endpoint}
\label{sec:SM_degenerate}

The eigenspace bookkeeping in this section is general. Whenever the lowest stiffness in a fixed symmetry sector has multiplicity greater than one, the complete lowest-stiffness eigenspace must be retained rather than selecting an arbitrary eigensolver vector. The Hilbert--Schmidt orthonormalization below is likewise a basis change within that retained space. The subsequent polar-decomposition search for a unitary combination is an additional selector for the clock endpoint problem, whose physical endpoint is unitary on the complete boundary Hilbert space. It is not an acceptance condition for the spin-one $D_2$ classifier, whose practical rule is given separately in Sec.~\ref{subsec:SM_spin1_final_rule}.

\subsection{Two different degeneracies}
\label{subsec:SM_two_degeneracies}

For each charge $q$, solve
\begin{equation}
D_qv=\kappa C_qv.
\label{eq:SM_sector_problem}
\end{equation}
There are two distinct cases. First, the lowest eigenvalue inside one fixed sector can have multiplicity $m_q>1$. This leaves the charge known and makes only the recovered operator ambiguous. Second, several charge sectors can share the global lowest stiffness. This leaves the charge itself ambiguous. The integer $m_q$ always means the dimension of the lowest-eigenvalue eigenspace inside sector $q$.

Let the columns of $V_q$ span this eigenspace and define temporary operators
\begin{equation}
\widetilde B_{q,i}=\sum_a(V_q)_{ai}O_{q,a},
\qquad
i=1,\ldots,m_q.
\label{eq:SM_degenerate_operators}
\end{equation}
The generalized eigenvectors are naturally orthonormal in the state metric, while unitarity is a matrix property measured by the normalized Hilbert--Schmidt product
\begin{equation}
\hs{A}{B}=\frac1{d_\partial}\Tr(A^\dagger B),
\qquad
d_\partial=d^\ell.
\label{eq:SM_HS_metric}
\end{equation}
Form the Hilbert--Schmidt Gram matrix $G_{ij}=\hs{\widetilde B_{q,i}}{\widetilde B_{q,j}}$ and diagonalize it as $G=S\Lambda S^\dagger$. After removing any zero eigenvalues, define
\begin{equation}
B_{q,i}=\sum_j\widetilde B_{q,j}(S\Lambda^{-1/2})_{ji}.
\label{eq:SM_HS_orthonormalize}
\end{equation}
Then $\hs{B_{q,i}}{B_{q,j}}=\delta_{ij}$. Apply the same right multiplication to the coefficient matrix, $V_q\leftarrow V_qS\Lambda^{-1/2}$, so each column of $V_q$ still gives the corresponding $B_{q,i}$ in the original operator basis. Every vector in an exactly degenerate eigenspace has the same $\kappa$, so this change of basis preserves the lowest stiffness.

Every normalized candidate is now
\begin{equation}
B_q(c)=\sum_{i=1}^{m_q}c_iB_{q,i},
\qquad
\sum_i|c_i|^2=1.
\label{eq:SM_Bc}
\end{equation}
It already has Hilbert--Schmidt norm one, as a unitary matrix does. If $m_q=1$, set $c_q^\star=1$. If $m_q>1$, the coefficients are chosen within this finite-dimensional sphere by the unitarity search below; $c_q^\star$ denotes the retained coefficient vector for sector $q$.

The connected metric intentionally discards scalar identity components. Indeed, from the definition $\delta A=A-\Tr(\rho A)\mathds 1$,
\begin{equation}
	\delta(A+z\mathds 1)=\delta A
\end{equation}
for any scalar $z$. Thus the variational matrices $C$ and $D$ cannot distinguish $A$ from $A+z\mathds 1$: they determine only the connected equivalence class of the operator. This causes no ambiguity for the traceless clock endpoints used in this work. It also cannot occur in a nontrivial charge sector, because the identity belongs to the neutral sector.

In a general application, however, a symmetry-neutral physical endpoint may have nonzero trace. Its connected representative can then appear nonunitary even though adding the omitted identity component would restore a unitary matrix. In that case the final unitarity search should use
\begin{equation}
	B_q(c)+z\mathds 1,
	\qquad
	z\in\mathbb C,
\end{equation}
and optimize over the free scalar $z$ together with the coefficients $c$. This restores exactly the identity component that was removed by connected subtraction; it does not affect the commutator because $[H,\mathds 1]=0$.

\subsection{Why singular-value decomposition is the correct local step}
\label{subsec:SM_polar_proof}

For a specified coefficient vector $c$, the matrix $B=B_q(c)$ is known. Compute its singular-value decomposition
\begin{equation}
B=W\Sigma Z^\dagger,
\qquad
\Sigma=\operatorname{diag}(\sigma_1,\ldots,\sigma_{d_\partial}),
\qquad
\sigma_j\geq0.
\label{eq:SM_SVD}
\end{equation}
A unitary matrix has every singular value equal to one. The matrix
\begin{equation}
Q=WZ^\dagger
\label{eq:SM_polar_Q}
\end{equation}
is a closest unitary matrix to $B$ in Frobenius norm.

\begin{lemma}[Closest-unitary property]
\label{lem:SM_closest_unitary}
For every unitary $U$,
\begin{equation}
\norm{B-Q}_{\rm F}^2=\sum_j(\sigma_j-1)^2\leq\norm{B-U}_{\rm F}^2.
\label{eq:SM_closest_unitary}
\end{equation}
\end{lemma}

\noindent\textit{Proof.} Expand
\begin{equation}
\norm{B-U}_{\rm F}^2=\norm{B}_{\rm F}^2+d_\partial-2\operatorname{Re}\Tr(U^\dagger B).
\label{eq:SM_Procrustes_expand}
\end{equation}
Writing $Y=W^\dagger UZ$, which is unitary, gives $\operatorname{Re}\Tr(U^\dagger B)=\operatorname{Re}\Tr(Y^\dagger\Sigma)\leq\sum_j\sigma_j$. Equality holds for $Y=\mathds 1$, equivalently $U=WZ^\dagger$. Substitution gives Eq.~\eqref{eq:SM_closest_unitary}. \hfill$\square$

Three useful diagnostics follow directly from the singular values:
\begin{equation}
	d_U(c)^2=\frac1{d_\partial}\sum_j(\sigma_j-1)^2,
	\qquad
	r_U(c)^2=\frac1{d_\partial}\sum_j(\sigma_j^2-1)^2.
	\label{eq:SM_unitarity_errors}
\end{equation}
The normalized Frobenius diagnostic is
\begin{equation}
	r_U(c)=
	\frac{\norm{B_q(c)^\dagger B_q(c)-\mathds 1}_{\rm F}}
	{\sqrt{d_\partial}},
	\label{eq:SM_unitarity_direct}
\end{equation}
while the acceptance residual is the operator-norm defect
\begin{equation}
	s_U(c)=
	\norm{B_q(c)^\dagger B_q(c)-\mathds 1}_{\rm op}
	=
	\max_j|\sigma_j^2-1|.
	\label{eq:SM_unitarity_spectral}
\end{equation}
All three quantities vanish exactly when $B_q(c)$ is unitary. The polar iteration uses $d_U$ to generate candidates and $r_U$ remains a useful averaged diagnostic, but acceptance uses $s_U$. In particular, every rank-deficient candidate has $s_U\geq1$, independently of $d_\partial$; a normalized Frobenius residual alone can instead become artificially small when the defect occupies a vanishing fraction of a growing Hilbert space.

Project the closest unitary back into the degenerate span:
\begin{equation}
z_i=\hs{B_{q,i}}{Q},
\qquad
c_i^{\rm new}=\frac{z_i}{\sqrt{\sum_j|z_j|^2}}.
\label{eq:SM_project_back}
\end{equation}
This formula is valid because the $B_{q,i}$ were Hilbert--Schmidt orthonormalized. Without that step, the projection coefficients are $G^{-1}z$ rather than $z$.

If $z=0$, this start has no projection back into the candidate space and is discarded. Otherwise repeat Eqs.~\eqref{eq:SM_SVD}, \eqref{eq:SM_polar_Q}, and \eqref{eq:SM_project_back}. Because an overall phase of a symmetry endpoint is physically irrelevant, convergence is measured by
\begin{equation}
\delta_c=\min_{\theta\in[0,2\pi)}\norm{c^{\rm new}-e^{\ii\theta}c}_2.
\label{eq:SM_phase_aligned_stop}
\end{equation}
Stop when both $\delta_c$ and the change in $d_U$ are below $10^{-12}$.

\begin{lemma}[What the alternating polar iteration guarantees]
\label{lem:SM_alternating}
Define
\begin{equation}
F(c,Q)=\operatorname{Re}\hs{Q}{B_q(c)},
\qquad
\norm{c}_2=1,
\qquad
Q^\dagger Q=\mathds 1.
\label{eq:SM_alternating_objective}
\end{equation}
The update $c\mapsto B_q(c)\mapsto Q\mapsto c^{\rm new}$ maximizes $F$ over $Q$ at fixed $c$ and then over $c$ at fixed $Q$. Hence $F$ never decreases, its values converge, and every convergent fixed point is stationary under the two block updates.
\end{lemma}

\noindent\textit{Proof.} Lemma~\ref{lem:SM_closest_unitary} shows that the polar factor maximizes $F$ over unitary $Q$. At fixed $Q$, write $F=\operatorname{Re}(z^\dagger c)$. Cauchy--Schwarz shows that its maximum on $\norm{c}_2=1$ is $\norm{z}_2$, attained by $c=z/\norm{z}_2$. The feasible sets are compact and $F$ is bounded, so the monotone objective values converge. \hfill$\square$

This lemma does not turn a nonconvex alternating iteration into a global optimizer. One initial vector can converge to a stationary point that is not the best combination. Multiple starts are a reproducible safety measure, while the final singular values provide the acceptance test.

\subsection{Deterministic starts and simple examples}
\label{subsec:SM_starts}

Let $e_i$ be the vector with one in entry $i$ and zero elsewhere. A compact deterministic start family is
\begin{equation}
\left\{e_i\right\}
\cup
\left\{\frac{e_i+e^{\ii\phi}e_j}{\sqrt2}:i<j,\ \phi\in\left\{0,\frac\pi2,\pi,\frac{3\pi}{2}\right\}\right\}
\cup
\left\{\frac{(1,\ldots,1)}{\sqrt{m_q}}\right\},
\label{eq:SM_start_family}
\end{equation}
with duplicates removed. Try $e_1$ first, then the other single directions, then pair mixtures, and finally the equal mixture. A run may stop inside a sector as soon as $s_U(c)\leq\tau_U$ supplies a positive unitary witness. This early stop does not claim that the retained residual is the sector minimum and must not be followed by ranking different passing sectors according to their subthreshold residuals. The single directions test whether an eigensolver already returned the endpoint. Pair mixtures test the simplest cancellations and the four elementary relative complex phases. The equal mixture tests cooperation among three or more directions. These starts are useful automation choices rather than a completeness theorem: failure of the finite list is inconclusive until a global search or bound excludes a unitary combination.

For $m_q=2$, the distinct starts are
\begin{equation}
(1,0),\ (0,1),\ \frac{(1,1)}{\sqrt2},\ \frac{(1,\ii)}{\sqrt2},\ \frac{(1,-1)}{\sqrt2},\ \frac{(1,-\ii)}{\sqrt2}.
\label{eq:SM_m2_starts}
\end{equation}
The reason that two degenerate directions can need more than two starts is that the answer may be a complex superposition rather than either basis vector. For the toy basis $B_1=\sqrt2|0\rangle\langle0|$ and $B_2=\sqrt2|1\rangle\langle1|$, each basis element is rank one, while $(B_1+B_2)/\sqrt2=\mathds 1_2$ is unitary.

For $m_q=3$, first use $(1,0,0)$, $(0,1,0)$, and $(0,0,1)$. If needed, use
\begin{equation}
\frac{(1,e^{\ii\phi},0)}{\sqrt2},
\qquad
\frac{(1,0,e^{\ii\phi})}{\sqrt2},
\qquad
\frac{(0,1,e^{\ii\phi})}{\sqrt2},
\qquad
\phi\in\left\{0,\frac\pi2,\pi,\frac{3\pi}{2}\right\},
\label{eq:SM_m3_pairs}
\end{equation}
and finish with $(1,1,1)/\sqrt3$. For $B_j=\sqrt3|j\rangle\langle j|$, the equal mixture is $\mathds 1_3$. More generally, coefficients $(1,e^{\ii\theta_1},e^{\ii\theta_2})/\sqrt3$ produce diagonal unitaries. For $m_q=4$, the same rule begins with four single directions, then uses four phases for each of the six pairs, and ends with $(1,1,1,1)/2$, for at most $4+24+1=29$ starts. In the explicit toy basis $B_j=2|j\rangle\langle j|$, $j=0,1,2,3$, the four basis matrices are rank one, whereas their equal mixture is $\mathds 1_4$.

\subsection{A rigorous global fallback}
\label{subsec:SM_global_fallback}

The unit sphere in $\mathbb C^{m_q}$ is compact, and $d_U(c)$ in Eq.~\eqref{eq:SM_unitarity_errors} is continuous. Therefore a global minimizer exists. There is also a simple finite certification principle. Because the basis is Hilbert--Schmidt orthonormal,
\begin{equation}
\norm{B_q(c)-B_q(c')}_{\rm F,n}=\norm{c-c'}_2.
\label{eq:SM_isometry_c}
\end{equation}
Distance to any closed set is one-Lipschitz, so
\begin{equation}
	|d_U(c)-d_U(c')|\leq\norm{c-c'}_2.
	\label{eq:SM_distance_Lipschitz}
\end{equation}
The operator-norm acceptance residual also has an explicit continuity bound. Since
$\norm{B_q(c)}_{\rm op}\leq\sqrt{d_\partial}$ and
$\norm{B_q(c)-B_q(c')}_{\rm op}\leq
\sqrt{d_\partial}\norm{c-c'}_2$,
\begin{equation}
	|s_U(c)-s_U(c')|
	\leq
	2d_\partial\norm{c-c'}_2.
	\label{eq:SM_spectral_Lipschitz}
\end{equation}
Evaluating $d_U$ on a finite $\delta$-net therefore bounds its global minimum within $\delta$, while evaluating $s_U$ bounds its global minimum within $2d_\partial\delta$. A branch-and-bound refinement can consequently certify unitary feasibility, or exclusion at the chosen tolerance, to any prescribed accuracy. This fallback grows rapidly with $m_q$, but it supplies the theorem-level guarantee that a finite list of polar starts cannot provide.

\subsection{An exactly solvable degeneracy: the $N=3,p=0$ fixed point}
\label{subsec:SM_p_zero_degeneracy}

Degenerate minima are not restricted to the trivial class. They can also occur at nonzero $p$, where the complete tied subspaces must be treated by the general procedure above. The case $N=3,p=0$ is useful because every accidental branch can be written explicitly. At $p=0$, the clock fixed point is unitarily equivalent to onsite terms
\begin{equation}
h=-\frac12(X_N+X_N^\dagger).
\label{eq:SM_p0_onsite}
\end{equation}
Let $|\widetilde r\rangle=N^{-1/2}\sum_{k=0}^{N-1}\omega_N^{-rk}|k\rangle$, so $X_N|\widetilde r\rangle=\omega_N^r|\widetilde r\rangle$. Then
\begin{equation}
h|\widetilde r\rangle=-\cos\left(\frac{2\pi r}{N}\right)|\widetilde r\rangle.
\label{eq:SM_p0_energies}
\end{equation}
The levels $r$ and $N-r$ are degenerate. Consequently
\begin{equation}
E_r=|\widetilde r\rangle\langle\widetilde{N-r}|
\label{eq:SM_rank_one_intertwiner}
\end{equation}
commutes with $h$ and transforms as
\begin{equation}
X_NE_rX_N^\dagger=\omega_N^{2r}E_r.
\label{eq:SM_intertwiner_charge}
\end{equation}
For $N=3$, $E_1$ and $E_1^\dagger$ therefore create exact zero-stiffness directions in the two nonzero charge sectors. After multiplication by the left anchor $X_0$, they remain rank deficient. Their singular values include zeros, so they cannot be unitary symmetry endpoints. The neutral candidate $X_0$ is unitary and gives $q_\star=0$.

For $N=4$, the excited states $r=1$ and $r=3$ are likewise degenerate and generate a charge-two zero-stiffness subspace. It acts only inside that two-dimensional excited subspace and remains rank deficient on the four-level site. This explains why degeneracy patterns depend on $N$ even though the correct trivial class is always $q_\star=0$.

The example isolates the physical issue: commutator minimization detects every operator acting inside an energy degeneracy, including rank-deficient maps and projectors. A symmetry endpoint is distinguished by its unitary action on the complete boundary Hilbert space, followed by the charge and edge-algebra checks. No prior knowledge of $p$ is used.

\subsection{Final decision rule for the clock endpoint search}
\label{subsec:SM_final_rule}

For each charge sector, retain the complete lowest-$\kappa$ eigenspace and its multiplicity $m_q$. Let $\kappa_{1,q}$ be its lowest stiffness, define $\kappa_{\min}=\min_q\kappa_{1,q}$, and let $\mathcal Q_{\min}$ contain every charge tied for this minimum within the chosen arithmetic tolerance. Because $\kappa_{\min}$ may vanish, this tolerance must include an absolute term scaled to the norm of the whitened eigenproblem; a purely relative tolerance proportional to $\kappa_{\min}$ is invalid. The charge is scanned over all sectors before any class is assigned.

For each $q\in\mathcal Q_{\min}$, use the Hilbert--Schmidt orthonormal basis $B_{q,i}$ defined in Eqs.~\eqref{eq:SM_HS_orthonormalize} and \eqref{eq:SM_Bc}. Here $d_\partial=d^\ell$ is the Hilbert-space dimension of the $\ell$-site boundary window; for the clock chain $d=N$, so $d_\partial=N^\ell$.

If $m_q=1$, there is only one normalized direction and it is tested directly. If $m_q>1$, process the deterministic starting vectors in Eq.~\eqref{eq:SM_start_family} one at a time. For each initial vector $c^{(0)}$, repeatedly apply Eqs.~\eqref{eq:SM_SVD}, \eqref{eq:SM_polar_Q}, and \eqref{eq:SM_project_back},
\begin{equation}
	c^{(0)}\longrightarrow c^{(1)}\longrightarrow c^{(2)}\longrightarrow\cdots,
\end{equation}
until the phase-aligned coefficient change in Eq.~\eqref{eq:SM_phase_aligned_stop} and the change in $d_U$ are below the chosen iteration tolerance. Denote the converged coefficient vector from that start by $c^{(\mathrm{run})}$ and evaluate
\begin{equation}
	r_U(c)=
	\frac{\norm{B_q(c)^\dagger B_q(c)-\mathds 1}_{\rm F}}
	{\sqrt{d_\partial}},
	\qquad
	s_U(c)=
	\norm{B_q(c)^\dagger B_q(c)-\mathds 1}_{\rm op}.
	\label{eq:SM_rq}
\end{equation}
Because the basis $B_{q,i}$ is Hilbert--Schmidt orthonormal and $\sum_i|c_i|^2=1$, $B_q(c)$ already has the same normalized Hilbert--Schmidt norm as a unitary matrix.

A converged run satisfying $s_U(c^{(\mathrm{run})})\leq\tau_U$ is a positive unitary witness. Retain its coefficients as $c_q^\star$ and stop trying further starts in that sector. If no run passes, complete the prescribed start list and retain the run with the smallest $s_U$ only for diagnostic purposes; failure of the finite list does not certify absence of a unitary combination. The reported choice $\tau_U=10^{-8}$ presumes IEEE complex double precision. It is not a reliable acceptance target in single precision, where machine epsilon already exceeds $10^{-8}$.

For the sector-level mathematical decision, define
\begin{equation}
	s_q^{\min}
	=
	\min_{\norm{c}_2=1}s_U(c),
	\qquad
	\mathcal A_\tau
	=
	\left\{
	q\in\mathcal Q_{\min}:s_q^{\min}\leq\tau_U
	\right\}.
	\label{eq:SM_qstar_final}
\end{equation}
The polar multi-start search supplies candidate upper bounds on $s_q^{\min}$ and proves membership in $\mathcal A_\tau$ whenever it finds a passing witness. Excluding a tied sector for a theorem-level classification requires the certified global fallback in Sec.~\ref{subsec:SM_global_fallback}; merely exhausting the finite start list is inconclusive. Accept a charge only when
\begin{equation}
	\mathcal A_\tau=\{q_\star\}.
\end{equation}
If $\mathcal A_\tau$ is empty, strengthen the search or enlarge the retained space; if it contains several sectors, the local classification is unresolved. Residuals that already lie below $\tau_U$ are not ranked against one another.

To reconstruct the physical endpoint, recall that the columns of $V_q$ introduced in Eq.~\eqref{eq:SM_degenerate_operators}, with the same Hilbert--Schmidt basis transformation applied after Eq.~\eqref{eq:SM_HS_orthonormalize}, contain the coefficients of the operators $B_{q,i}$ in the original anchored basis $\{O_{q,a}\}$. Therefore
\begin{equation}
	v_{\rm rec}=V_{q_\star}c_{q_\star}^\star
	\label{eq:SM_recovered_coefficients}
\end{equation}
converts the coefficients $c_{q_\star}^\star$ from the retained degenerate eigenspace into coefficients $v_{\rm rec}$ in the original boundary-operator basis. The recovered left endpoint is then
\begin{equation}
	A_1^\ell
	=
	B_{q_\star}(c_{q_\star}^\star)
	=
	\sum_a(v_{\rm rec})_aO_{q_\star,a},
	\qquad
	p_{\rm out}=-q_\star\pmod N.
	\label{eq:SM_final_recovery}
\end{equation}
The last relation follows from Eq.~\eqref{eq:SM_p_readout}: an endpoint of charge $q_\star$ corresponds to the projective class $p=-q_\star\pmod N$. Thus the final sequence is
\begin{equation}
	q_\star
	\longrightarrow
	c_{q_\star}^\star
	\longrightarrow
	v_{\rm rec}=V_{q_\star}c_{q_\star}^\star
	\longrightarrow
	A_1^\ell
	\longrightarrow
	p_{\rm out}.
\end{equation}
In the ordinary nondegenerate case, $m_q=1$ and $c_q^\star=1$, so $V_q$ has only the single retained eigenvector column and the procedure directly reconstructs that normalized eigenoperator. Accept a class only when the selected charge sector contains a near-unitary endpoint, its residual in Eq.~\eqref{eq:SM_charge_algebra_residuals} is controlled, the branch is stable under increasing $\ell$, its open-chain stiffness is boundary separated according to Eq.~\eqref{eq:SM_eta}, and the independently recovered $A_2^\ell$ satisfies the projective algebra. If the unique lowest branch is rank deficient or otherwise far from unitary, it is not an endpoint: enlarge the retained low-stiffness band to include the next eigenspaces and repeat the unitarity search before increasing the boundary window. If no low-stiffness sector contains a near-unitary candidate, enlarge the window or anchored space. If several charge sectors contain equally admissible unitary candidates after the global check, the local classification is unresolved and requires explicit global-symmetry factorization or an independent ground-state projective-representation calculation. Sector order and eigensolver order carry no physical information.

\subsection{Practical decision rule for spin-one $D_2$ classification}
\label{subsec:SM_spin1_final_rule}

The spin-one calculation uses the same generalized eigenproblem but a different final readout. Most importantly, classification is entered only after detection has established protected boundary memory using the depth and size flow, open-versus-periodic and boundary-versus-bulk comparisons, and symmetry-preserving boundary perturbations. The generalized eigenproblem may still be evaluated when detection fails, as for the large-$D$ control in the Letter, but its eigenvalues are then reported only as variational stiffnesses and no protected projective class is assigned.

Once detection has passed, choose a boundary window $\ell$ for which the boundary branch is stable and solve Eq.~\eqref{eq:SM_generalized_problem} independently in all three nontrivial $D_2$ sectors,
\begin{equation}
	(0,1),\qquad(1,0),\qquad(1,1).
	\label{eq:SM_spin1_practical_sectors}
\end{equation}
The sectors $(0,1)$ and $(1,0)$ provide the two independent operators required for the class readout,
\begin{equation}
	A_x^\ell\in(0,1),
	\qquad
	A_z^\ell\in(1,0),
	\label{eq:SM_spin1_required_pair}
\end{equation}
while the $(1,1)$ operator $A_y^\ell$ is an independent redundancy check. One nontrivial sector alone is therefore not sufficient to determine the $D_2$ projective class.

For each sector, retain the complete lowest-stiffness eigenspace and record its multiplicity $m_{\bm q}$. If $m_{\bm q}=1$, the normalized lowest generalized eigenvector directly gives the recovered operator in that sector. If $m_{\bm q}>1$, retain the complete degenerate space and use the Hilbert--Schmidt orthonormalization and window-stability bookkeeping described above. The physical class must not depend on which arbitrary basis the eigensolver returns inside that degenerate space. If a particular representative is needed from a degenerate $A_x$ or $A_z$ space, write
\begin{equation}
	A_x(c_x)=\sum_{i=1}^{m_{(0,1)}}(c_x)_iB_{(0,1),i},
	\qquad
	A_z(c_z)=\sum_{j=1}^{m_{(1,0)}}(c_z)_jB_{(1,0),j},
	\qquad
	\norm{c_x}_2=\norm{c_z}_2=1,
	\label{eq:SM_spin1_degenerate_pair}
\end{equation}
and choose combinations that minimize the endpoint-algebra residual. The full-window polar unitarity residual used in the clock decision rule is not imposed here.

The spin-one analogue of the clock label is the sign in the algebra of the two independently recovered endpoint operators. Because these endpoints are constrained by the stiffness only on the thermally weighted sector, the residual is measured in the state norm $\norm{M}_S=\sqrt{(M|M)_S}$ of Eqs.~\eqref{eq:SM_inner} and \eqref{eq:SM_norm}, with $A_x(c_x)$ and $A_z(c_z)$ rescaled to $\norm{A_x}_S=\norm{A_z}_S=1$, rather than in the normalized Frobenius norm: compare
\begin{equation}
	r_{D_2}^{(p)}
	=
	\min_{c_x,c_z}
	\norm{
		A_x(c_x)A_z(c_z)-(-1)^pA_z(c_z)A_x(c_x)
	}_S,
	\qquad
	p\in\{0,1\},
	\label{eq:SM_spin1_algebra_residual}
\end{equation}
where the minimization is absent when both sectors are nondegenerate. The projective label is
\begin{equation}
	A_x^\ell A_z^\ell\simeq(-1)^{p_{D_2}}A_z^\ell A_x^\ell,
	\qquad
	p_{D_2}\in\{0,1\}.
	\label{eq:SM_spin1_class}
\end{equation}
The value $p_{D_2}=1$ is the nontrivial Haldane class and corresponds to anticommuting endpoint actions, while $p_{D_2}=0$ denotes the commuting trivial projective class. Under the staged protocol used here, however, a model that fails detection is not assigned $p_{D_2}=0$ merely because its generalized eigenproblem can be solved; the large-$D$ chain is therefore a no-detection control rather than a forced trivial-class output.

Finally, use the $(1,1)$ sector as a redundancy check. For the nontrivial class, the recovered $A_y^\ell$ should be consistent with the same effective Pauli algebra, so its state-norm anticommutator residuals with $A_x^\ell$ and $A_z^\ell$, defined as in Eq.~\eqref{eq:SM_spin1_algebra_residual}, should also be small; at fixed finite $\beta$ they saturate at a small floor set by thermal excitations inside the window rather than vanishing, and vanish only as $\beta\to\infty$. For isotropic AKLT, rotational symmetry additionally makes the three nontrivial stiffness branches equivalent, which is why the End Matter displays one representative AKLT sector.

The two practical readouts can therefore be summarized as
\begin{equation}
	\boxed{
		\begin{aligned}
			\mathbb Z_N\times\mathbb Z_N:\quad&
			\text{detection}
			\longrightarrow
			\text{anchored charge scan}
			\longrightarrow
			q_\star
			\longrightarrow
			p=-q_\star\pmod N,\\
			D_2:\quad&
			\text{detection}
			\longrightarrow
			\{(0,1),(1,0),(1,1)\}
			\longrightarrow
			(A_x,A_z;\ A_y\text{ check})
			\longrightarrow
			p_{D_2}\text{ from }A_xA_z\simeq(-1)^{p_{D_2}}A_zA_x.
	\end{aligned}}
	\label{eq:SM_classification_comparison}
\end{equation}

The complete logical chain is
\begin{equation}
\boxed{
\begin{gathered}
\text{Gibbs-weighted metric and Hermitian initial operator}
\Longrightarrow
\text{even frequency weight and zero-diagonal Krylov chain},\\
\{b_n\}
\Longrightarrow
\Zz_K,\ A_K,\ \varepsilon_K
\Longrightarrow
\text{exact or finite-depth memory under their stated limits},\\
\text{complete boundary space, charge projection, and endpoint algebra}
\Longrightarrow
\text{bosonic projective class}.
\end{gathered}}
\label{eq:SM_dependency_chain}
\end{equation}
Each implication uses the assumptions stated in its section. Detection establishes boundary memory after geometry and symmetry-preserving boundary-perturbation controls, and classification is entered only after those tests have passed. Classification then identifies the projective symmetry action using the model-appropriate final readout: the anchored charge and full-window unitary endpoint test for the clock models, or the algebra of independently recovered nontrivial $D_2$ sectors for the spin-one problem. For the clock endpoint search, the full-window unitarity test in Sec.~\ref{subsec:SM_final_rule} is an additional acceptance condition; for spin-one $D_2$, the projective readout is instead the algebra of the independently recovered nontrivial sectors described in Sec.~\ref{subsec:SM_spin1_final_rule}.

\bibliography{refs_kr}